\pdfoutput=1
\documentclass[twoside, 12pt]{iiser-thesis}

\usepackage[american]{babel}
\usepackage{fullpage}
\usepackage{amssymb,latexsym,amsmath}     
\usepackage{graphicx}
\usepackage{authblk}
\usepackage{dsfont}
\usepackage{amsthm}
\usepackage{color}
\usepackage{float}
\usepackage{svg}
\usepackage{wrapfig}
\usepackage{pdfpages}
\usepackage{hyperref}
\hypersetup{
colorlinks = true,
allcolors=blue
}
\usepackage[backend=biber, bibencoding=utf8,citestyle=numeric, style=ieee, doi=false, url=true, eprint=false]{biblatex}
\AtEveryBibitem{
 \clearlist{address}
 \clearfield{date}
 \clearfield{eprint}
 \clearfield{isbn}
 \clearfield{issn}
 \clearlist{location}
 \clearfield{month}
 \clearfield{series}
 \clearfield{issue}
}
\graphicspath{{images/}}
\usepackage{tikz}
\usetikzlibrary{matrix,shapes,arrows,positioning,chains,arrows.meta}
\catcode`\@=11
\catcode`\@=12

\title{\textbf{Hamiltonian Engineering in Quantum Spin Networks}}

\author{\textbf{Ieshan Vaidya}}
\coordinator{Coordinator} \supervisor{Dr. T. S. Mahesh} \sdesignation{Associate Professor}
\department{Department of Physics} \reader{Dr. Rejish Nath}
\dedication{This thesis is dedicated to my family.}
\graduationyear{2018}
\academicyear{2017-2018}
\graduationmonth{May}

\thesisabstract{Quantum simulation presents itself as one of the biggest advantages of developing quantum computers. Simulating a quantum system classically is almost impossible beyond a certain system size whereas a controllable quantum system inherently has the resources and computing space to simulate another system. Analog quantum simulation is one of the ways of quantum simulation through which a known system mimics an unknown system. A key aspect of this is the ability to generate the target Hamiltonian using control operations which is referred to as Hamiltonian engineering. One way of doing this is to apply pulse sequences over a length of time such that the average Hamiltonian over this period is the desired one. In this thesis, we discuss the method of filtered Hamiltonian engineering which works in a similar fashion. Using this technique, we create a star topology from a general network of spins. 

Quantum communication between two parties is an important task for its applications in information theory as well as for its use in a quantum computer. A typical solution to this is using teleportation through the means of shared entangled qubits. Teleportation is not ideal for short-range communication such as between two units in a quantum computer. It has been shown that information can be transported from one node of a quantum spin network to another by natural evolution of the system over time. Such networks are better suited for solid-state based computing architectures as well as for short-range communication. The Hamiltonian that permits such transport however places stringent requirements on the parameters of the network. While individual control of spins cannot be avoided in most cases, excess control can introduce noise. In this thesis, we work on a few models of spin networks that permit information transport with minimal requirements on the parameters of the network.
}
\acknowledgments{The past 5 years leading up to this thesis have been memorable. First and foremost, I express my sincere gratitude to my supervisor, Professor T. S. Mahesh for his guidance, support and resources during the past two and a half years. Whilst working with him, I learned a lot about quantum information science as well as research work in general. The discussions and group-meetings with him and fellow group-mates Deepak Khurana, Anjusha V. S., V. R. Krithika, Soham Pal, C S Sudheer Kumar and Gaurav Bhole were insightful and enjoyable. I am especially thankful to Soham for his inputs on the topic of routing quantum information. I am grateful to Professor Rejish Nath for introducing me to the world of quantum physics. I worked with him for a brief time and learned a lot about theoretical physics as well as computational techniques. I am thankful for his inputs on my thesis work as a member of my thesis advisory committee. I would also like to thank my family and friends for their motivation and support. I would like to express my gratitude towards the DST-INSPIRE program under the Department of Science and Technology,  Government of India for providing financial support through my study at IISER Pune.}

\begin{document}
	\thesisfront
	\listoffigures
	\listoftables

\tikzset{
desicion/.style={
    diamond,
    draw,
    text width=4em,
    text badly centered,
    inner sep=0pt
},
block/.style={
    rectangle,
    draw,
    text width=10em,
    text centered,
    rounded corners
},
cloud/.style={
    draw,
    ellipse,
    minimum height=2em
},
descr/.style={
    fill=white,
    inner sep=2.5pt
},
connector/.style={
    -latex,
    font=\scriptsize
},
rectangle connector/.style={
    connector,
    to path={(\tikztostart) -- ++(#1,0pt) \tikztonodes |- (\tikztotarget) },
    pos=0.5
},
rectangle connector/.default=-2cm,
straight connector/.style={
    connector,
    to path=--(\tikztotarget) \tikztonodes
}
}

\chapter*{Units, Definitions and Notations}
\label{chap:defs}
\begin{enumerate}
\item In all discussions, we assume $\hbar = 1$. Unit of time is seconds.
\item \textbf{Fidelity} \\
It is a measure of distance between two density matrices $\rho$ and $\sigma$ defined as \cite{NielsenChuang}
$$ F\left(\rho,\sigma\right)=tr\sqrt{\rho^{\frac{1}{2}}\sigma\rho^{\frac{1}{2}}} $$
\item \textbf{Spin Commutation Relations} \\
General commutation relation between spin operators is given by
$$ \left[S_i^p,S_j^q\right] = i\delta_{ij}\epsilon_{pqr}S_i^r$$
where $i,j$ are spatial designations; $p,q,r\in(x,y,z)$; $\delta_{ij}$ is the Kronecker delta and $\epsilon_{pqr}$ is the Levi-Civita symbol.
\item \textbf{Dipolar Hamiltonian}
$$\mathcal{H}_D = \sum_{i,j} b_{ij}\left(3S_i^zS_j^z-S_i \cdot S_j\right)$$
\item \textbf{Double-Quantum Hamiltonian}
$$\mathcal{H}_{DQ} = \sum_{i,j} b_{ij}\left(S_i^xS_j^x-S_i^yS_j^y\right)$$
\item \textbf{XY Hamiltonian}
$$\mathcal{H}_{XY} = \sum_{i,j} b_{ij}\left(S_i^xS_j^x+S_i^yS_j^y\right)$$
\end{enumerate}
	
\chapter{Introduction}
Quantum computation \cite{NielsenChuang} has achieved remarkable progress in recent years. Notable advances have been made in the design of quantum algorithms, quantum error-correcting codes, quantum cryptography, quantum communication as well as in the realization of experimental architectures. Quantum computers provide more efficient solutions to some problems than a classical computer. One of the biggest uses of a quantum computer is in the area of quantum simulation \cite{QSimReview}. Simulation of a quantum system on a classical computer is computationally hard. The memory required to encode a quantum system on a classical computer grows exponentially with the input. The operators that determine the evolution of the system also grow exponentially and consequently simulating a quantum system beyond tens of qubits becomes intractable. In 1981, Richard Feynman envisioned using known quantum systems to simulate other quantum systems \cite{Feynman1982} since they inherently capture the extra computing space that is classically unavailable. Building on this idea, it was shown that a quantum computer can act as a universal quantum simulator \cite{LloydUniversalQSim}. This approach of using unitary gates to create the target propagator is called digital quantum simulation. The parallel to this is analog quantum simulation where a known quantum system is used to mimic the target one. A third approach is through adiabatic means where we start with the ground state of a known system and adiabatically move to the target system \cite{AdiabaticQC}. The field of quantum simulation has grown rapidly and promises applications in diverse areas of Physics \cite{ApplCMP3, ApplCMP5, ApplHEP2,
ApplHEP5,ApplC2,ApplAP1,ApplOQS2,ApplOQS6,ApplQC1,ApplQC2,ApplI1} as well as in Chemistry \cite{ApplChem1,ApplChem2,ApplChem4,
ApplChem5,ApplChem6} and Biology \cite{ApplBio1,ApplBio2}. 

Quantum simulation requires the control of a known quantum system whose Hamiltonian parameters can be altered such that it behaves like an unknown system we wish to simulate. Broadly, quantum simulation involves the following steps \cite{QSimGoals} :
\begin{enumerate}
\item Initialization to a known state
\item Engineering the desired Hamiltonian
\item Detecting and verifying the required state
\end{enumerate}

In this study, we focus on the area of Hamiltonian engineering. It refers to engineering the parameters of a desired Hamiltonian by applying control operations on a known Hamiltonian \cite{HamEngSonia}. We employ the approach of filtered Hamiltonian engineering proposed by Ajoy, Cappellaro \cite{AjoyHamEng}. In this approach, a network of spins is allowed to evolve alternately under its internal Hamiltonian and a Zeeman field Hamiltonian. We tune the parameters such that the total unitary after the entire sequence is on average the required propagator. Spin networks are a graph of spin $\frac{1}{2}$ particles interconnected with each other with some strengths. Such networks has practical applications in quantum simulation \cite{SpinNetQSim}, quantum communication \cite{BoseQComm} and are also useful in theoretical studies such as in studying decision tree problems \cite{FarhiDecTree}. A particularly useful topology of spin networks is the star topology. A network in star topology has one central spin connected to other peripheral spins. Star topology is routinely seen in classical computing in the form of hubs or registers; along similar lines, there are many applications in the quantum domain such as information routing \cite{KayRouting,SpinStarSwitch}, state amplification \cite{KayStateAmpli} and quantum sensing \cite{MagSensing} among others. Finding a system naturally present in a star topology is rare and might still have weak interaction among other nodes. It is desirable to obtain a perfect star topology from a network of spins. In this study, we propose a filtered approach to decouple all the unwanted interactions and retain the necessary interactions to obtain a star topology.

Sharing of quantum resources across two spatially separated parties requires a quantum communication channel. Quantum communication is crucial for the development of a quantum computer as well as for quantum cryptography. Sharing of information can be done by transmitting the state directly or by teleportation \cite{QTeleport1,QTeleport2}. For a spin based system, this would typically require encoding the information in an optical channel which is not ideal. Additionally, it's suboptimal for short range communication. Bose showed that it is possible to transport information from one node of a quantum spin network to another by the natural evolution of the network \cite{BoseQComm}. Numerous schemes \cite{QTSchemes1,QTSchemes2,QTSchemes3,QTSchemes4,
QTSchemes5,QTSchemes6} have been proposed for state transport in quantum networks. The limitations of using spin networks for state transfer lie in the realization of the Hamiltonian. Enabling state transport requires specific interaction strengths that would typically require local control on the spins which is difficult and additionally introduces noise. 

In this study, we propose a few models of information transport in spin networks that have minimal requirements of the Hamiltonian parameters. We first consider a simple spin chain that permits state transport from one end to another. We assume that the chain is uniformly coupled and there is a Zeeman field only on the ends of the chain. This creates a resonance effect in the chain where the ends of the chain talk to each other and exchange information. We further extend this resonance effect to a routing mechanism where it is possible to send information from the input port to one of the two output ports based on a control condition. Many routing protocols have been proposed \cite{Routing1,Routing2,Routing3,Routing4} before. We propose a simple protocol which requires only a switch of the magnetic field on the input port and thus is non-invasive. We consider two models of spin systems that achieve this. We show a simple extension of the spin chain to a 4-spin system with a central node that acts as the routing node connected to the input node and two output nodes. Additionally, we show routing in a 5-spin system based on the concept of conditional state transfer using the two central spins acting as a controlling gate \cite{SpinTransistor}. While such transport models can be extended to longer lengths, they typically lose their accuracy with size and the time required to transport information also increases. Although one can decrease the required time by amplifying the parameters, this places strong requirements on the system. A possible alternative is to combine two blocks of spins so that they achieve the combined purpose of the two individual blocks. We discuss such a scheme where two separate information transport blocks can be combined that serve the purpose of the individual blocks in a modular fashion. We show that this is possible by using time-dependent magnetic fields on specific spins. This can also be extended to multiple blocks, however there is a small drop in accuracy with each additional module. To complete the discussion on information transport in spin networks, we consider a general spin network where we comment on the resonance effect and show that transport can still be achieved albeit with lower accuracy. 

The thesis work can be summarized by the diagram in Figure \ref{summary} below.

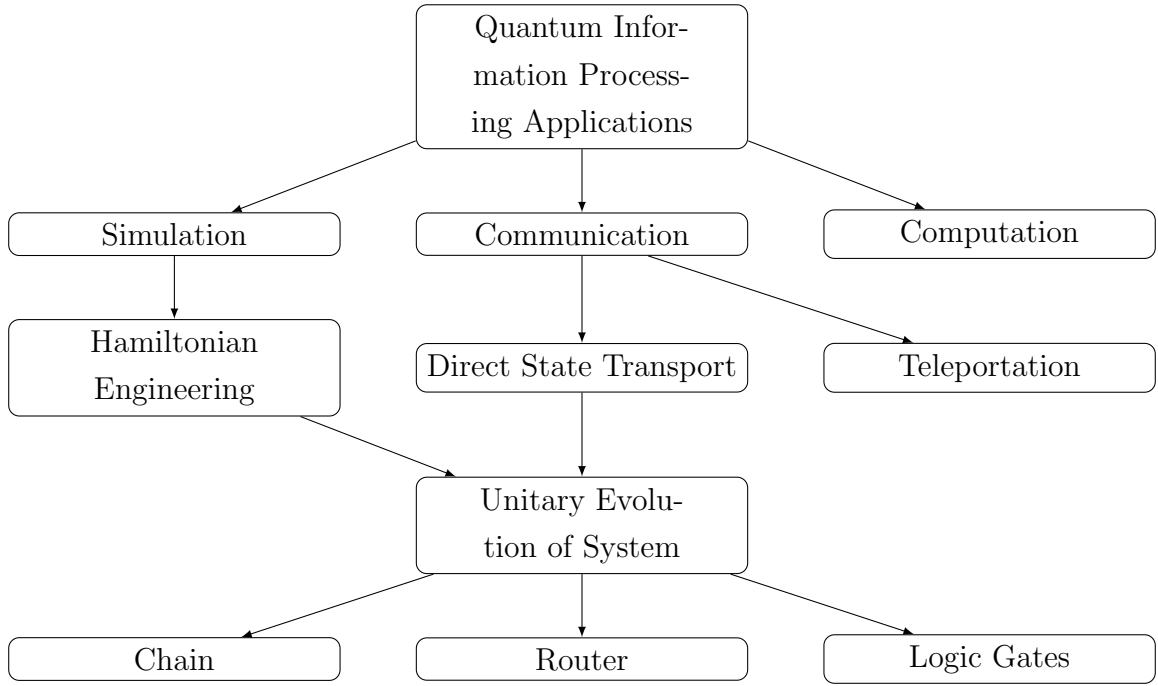
\begin{figure}[H]
\centering
\begin{tikzpicture}
\matrix (m)[matrix of nodes, column  sep=1cm,row  sep=8mm, align=center, nodes={rectangle,draw, anchor=center} ]{
      & |[block]| {Quantum Information Processing Applications} & \\
      |[block]| {Simulation} & |[block]| {Communication} & |[block]| {Computation}\\
      |[block]| {Hamiltonian Engineering}& |[block]| {Direct State Transport} & |[block]| {Teleportation} \\
      & |[block]| {Unitary Evolution of System} & \\
      |[block]| {Chain} & |[block]| {Router} & |[block]| {Logic Gates} \\
};
\path [>=latex,->] (m-1-2) edge (m-2-1);
\path [>=latex,->] (m-1-2) edge (m-2-2);
\path [>=latex,->] (m-1-2) edge (m-2-3);
\path [>=latex,->] (m-2-1) edge (m-3-1);
\path [>=latex,->] (m-2-2) edge (m-3-2);
\path [>=latex,->] (m-2-2) edge (m-3-3);
\path [>=latex,->] (m-3-1) edge (m-4-2);
\path [>=latex,->] (m-3-2) edge (m-4-2);
\path [>=latex,->] (m-4-2) edge (m-5-1);
\path [>=latex,->] (m-4-2) edge (m-5-2);
\path [>=latex,->] (m-4-2) edge (m-5-3);
\end{tikzpicture}
\caption{Summary of Thesis Work}
\label{summary}
\end{figure}

\chapter{Star Topology Engineering}
\begin{wrapfigure}{r}{0.3\linewidth}
	\centering
            \def\svgwidth{0.8\linewidth}
		    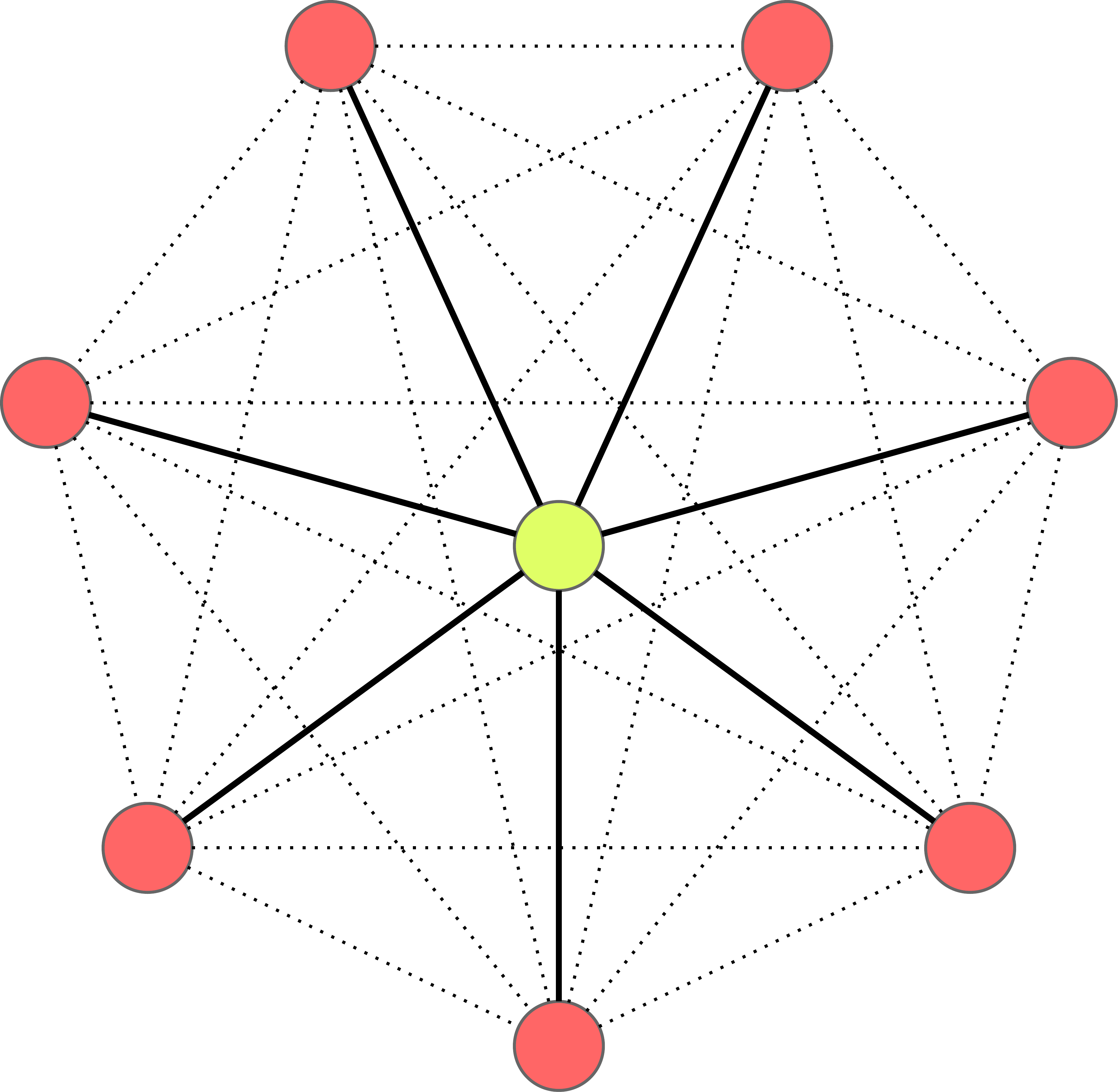
		    \caption{Star Topology in a Network with 8 Spins}
		    \label{fig:img_star}
\end{wrapfigure}
Filtered Hamiltonian engineering refers to using multiple unitary sequences whose average effect is the target operator. Ajoy and Cappellaro \cite{AjoyHamEng} used this mechanism to construct a Hamiltonian that permits state transport in a linear chain of spins. The term \textsc{\char13}filter\textsc{\char13} appears due to the role of a filter or grating function that retains only specific interactions and thus enables the engineering of a specific Hamiltonian. We apply to this technique to obtain a star topology of spins. This technique works even if every spin is connected to every other spin. A star topology is shown in Figure \ref{fig:img_star} where the dashed lines indicate the decoupled interactions and the solid lines show the retained interactions.

\section{System}
The system is a network of $n$ spin $\frac{1}{2}$ particles with one central spin and other indistinguishable peripheral spins. All the spins are assumed to be connected to one another with some strength. The objective is to decouple all the peripheral-peripheral interactions and retain the central-peripheral interactions using a filtered technique. The system is governed by the Double-Quantum (DQ) Hamiltonian along with Zeeman terms. We consider the DQ Hamiltonian for the system since it can be obtained from the dipolar Hamiltonian by pulse sequences \cite{DQPulseSeq}.   
  \begin{flalign}
  \mathcal{H}_{DQ} =& \sum_{i<j} b_{ij} \left(S_i^xS_j^x - S_i^yS_j^y\right) \\
  \mathcal{H}_{Z} =& \sum_{i} \omega_i S_i^z &&
  \end{flalign}
  The filtered scheme consists of alternate evolution of the system under the DQ Hamiltonian and the Zeeman Hamiltonian. During the DQ evolution, the Zeeman evolution is switched off and similarly during the Zeeman evolution, the DQ evolution is switched off. One can obtain only the Zeeman Hamiltonian by decoupling techniques \cite{LevittSD} and similarly can obtain only the DQ Hamiltonian by switching off the magnetic field. In subsequent discussions, at any point of time, we assume that the system is governed by either $\mathcal{H}_{DQ}$ or $\mathcal{H}_z$ only.

\section{Filtered Hamiltonian Engineering}
\begin{figure}[H]
	\centering
            \def\svgwidth{0.5\linewidth}
		    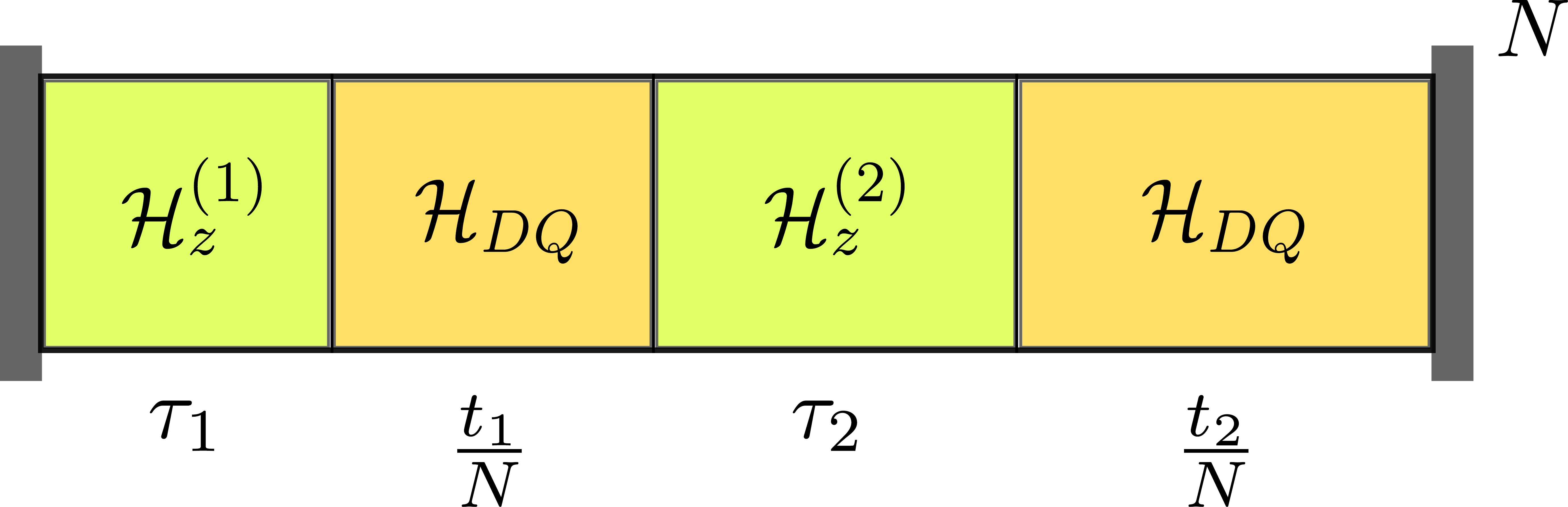
            \caption{Filtered Scheme for $L=2$}		    
		    \label{fig:img_filter}
\end{figure}

The filtered engineering scheme consists of alternate evolutions of the system under the DQ Hamiltonian and Zeeman Hamiltonian. We  call one alternate evolution as a \textsc{Stage} and \textbf{L} such stages as one \textsc{Sequence}. The entire scheme has \textbf{N} such sequences. The Zeeman evolutions have time period $\left(\tau_1, \tau_2, \cdots, \tau_L\right)$ and the DQ evolutions have time period $\left(\frac{t_1}{N}, \frac{t_2}{N}, \cdots, \frac{t_L}{N}\right)$. As a simple visualization, filtered scheme for $L=2$ is shown in Figure \ref{fig:img_filter} (In the figure, blocks are applied from the left; $\mathcal{H}_z^{(1)}$ is applied first and so on).
\noindent		 	 
At stage $i$, the Zeeman frequency of the central spin is $\Omega_i$. We have assumed that the peripheral spins are indistinguishable and so the magnetic fields on them at any stage must be identical. Consequently, the Zeeman frequency of the peripheral spins at stage $i$ is $\omega_i$. Thus, the Zeeman Hamiltonian at stage $i$ of a sequence is given by
		 \begin{flalign}
  \mathcal{H}_z^{(i)} = \omega_i \sum_{j=2}^n S_j^z + \Omega_iS_1^z &&
  \end{flalign}
The propagator for a 2-stage scheme is given by (setting $\tau_1 = \tau_2 = \tau$ for simplicity)
\begin{flalign}
		\label{totprop}
       U_N = \left[ U_{DQ}\left(\frac{t_2}{N}\right) U_Z^{(2)}\left(\tau\right) U_{DQ}\left(\frac{t_1}{N}\right) U_Z^{(1)}\left(\tau\right) \right]^N &&
       \end{flalign}
       We can introduce identity operators in the form of $U^{\dagger}U$ to obtain terms of similar structure. We can thus write the above equation as 
        \begin{flalign}
        \begin{split}
            U_N =& \left[U_Z^{(1)}\left(\tau\right)^\dagger U_Z^{(2)}\left(\tau\right)^\dagger \cdots U_{DQ}\left(\frac{t_2}{N}\right) U_Z^{(1)}\left(\tau\right) U_Z^{(2)}\left(\tau\right)\cdots \right]\cdot \\
            & \qquad \qquad \qquad \vdots \\
            & \left[U_Z^{(1)}\left(\tau\right)^\dagger U_Z^{(2)}\left(\tau\right)^\dagger U_Z^{(1)}\left(\tau\right)^\dagger U_{DQ}\left(\frac{t_1}{N}\right)  U_Z^{(1)}\left(\tau\right) U_Z^{(2)}\left(\tau\right) U_Z^{(1)}\left(\tau\right)\right]\cdot \\
            & \left[U_Z^{(1)}\left(\tau\right)^\dagger U_Z^{(2)}\left(\tau\right)^\dagger U_{DQ}\left(\frac{t_2}{N}\right) U_Z^{(2)}\left(\tau\right) U_Z^{(1)}\left(\tau\right) \right]\cdot \\
            & \left[U_Z^{(1)}\left(\tau\right)^\dagger U_{DQ}\left(\frac{t_1}{N}\right) U_Z^{(1)}\left(\tau\right)\right]
            \end{split}   &&
        \end{flalign}
        We will simplify each of the bracket-terms so that they have the same structure. To do so, we can write (since $\mathcal{H}_Z^{(1)} $ and $\mathcal{H}_Z^{(2)} $ commute with each other)
        \begin{flalign}
        \begin{split}
            U_Z^{(2)}\left(\tau\right) U_Z^{(1)}\left(\tau\right) =& \text{ exp}\left[-i \left(\mathcal{H}_Z^{(2)} + \mathcal{H}_Z^{(1)}\right)\tau\right] \\
            =& \text{ exp}\left[{-i \left(\omega_2 \sum_{j=2}^5 S_j^z + \Omega_2 S_1^z + \omega_1 \sum_{j=2}^5 S_j^z + \Omega_1 S_1^z\right)\tau}\right] \\
            =& \text{ exp}\left[{-i \left(\left(\omega_1 + \omega_2\right) \sum_{j=2}^5 S_j^z + \left(\Omega_1 + \Omega_2\right) S_1^z\right)\tau}\right] \\
            =& \; U_Z^{(1_1 2_1)}\left(\tau\right)
        \end{split} &&
        \end{flalign}
        The notation $(1_p2_q)$ implies that there is a multiplicative coefficient of $p$ on the stage one Zeeman frequency and there is a multiplicative coefficient of $q$ on the stage two Zeeman frequency. Concretely,
        \begin{flalign}
        U_Z^{(1_p 2_q)}\left(\tau\right) = \text{ exp}\left[{-i \left(\left(p\omega_1 + q\omega_2\right) \sum_{j=2}^5 S_j^z + \left(p\Omega_1 + q\Omega_2\right) S_1^z\right)\tau}\right]     
        &&
        \end{flalign}
        Similarly,
        \begin{flalign}
        \begin{split}
            U_Z^{(1)}\left(\tau\right) U_Z^{(2)}\left(\tau\right) U_Z^{(1)}\left(\tau\right) =& \text{ exp}\left[{-i \left(\left(2\omega_1 + \omega_2\right) \sum_{j=2}^5 S_j^z + \left(2\Omega_1 + \Omega_2\right) S_1^z\right)\tau}\right] \\
            =& \; U_Z^{(1_2 2_1)}\left(\tau\right)
        \end{split} &&
        \end{flalign}
        The total propagator now reads as
		 \begin{flalign}
            \begin{split}
            U_N =& \left[U_Z^{(1_N 2_N)}\left(\tau\right)^\dagger U_{DQ}\left(\frac{t_2}{N}\right)  U_Z^{(1_N 2_N)}\left(\tau\right)\right] \\
            & \qquad \qquad \qquad \vdots \\
            & \left[U_Z^{(1_2 2_1)}\left(\tau\right)^\dagger U_{DQ}\left(\frac{t_1}{N}\right) U_Z^{(1_2 2_1)}\left(\tau\right)\right]\cdot \\
            & \left[U_Z^{(1_1 2_1)}\left(\tau\right)^\dagger U_{DQ}\left(\frac{t_2}{N}\right) U_Z^{(1_1 2_1)}\left(\tau\right)\right]\cdot \\
            & \left[U_Z^{(1_1 2_0)}\left(\tau\right)^\dagger U_{DQ}\left(\frac{t_1}{N}\right) U_Z^{(1_1 2_0)}\left(\tau\right)\right]
            \end{split} &&
        \end{flalign}
        Each individual bracket has a similar form given by
        \begin{flalign}
        U_Z\left(\tau\right)^\dagger U_{DQ}\left(t\right) U_Z\left(\tau\right) = \text{ exp}\left[{-i t \mathcal{H}_m\left(\tau\right)}\right] &&
        \end{flalign}
        where $\mathcal{H}_m\left(\tau\right)$ is the Toggling Frame Hamiltonian (See Appendix \ref{appendix:tfh}) given by
        \begin{flalign}
        \label{eqn:tfh}
        \mathcal{H}_m\left(\tau\right) = \sum \frac{b_{ij}}{2} \left(S_i^+ S_j^+ e^{-i\tau \delta_{ij}} + S_i^- S_j^- e^{i\tau \delta_{ij}}\right) &&
        \end{flalign}
        and $\delta_{ij} = \omega_i + \omega_j$ is the sum of the frequencies on spin i and spin j.
        In terms of $\mathcal{H}_m$, the total propagator is
        \begin{flalign}
        \begin{split}
            U_N = & \text{ exp}\left[-i \frac{t_2}{N} \mathcal{H}_m^{(1_N 2_N)}\right] \cdot \text{ exp}\left[-i \frac{t_1}{N} \mathcal{H}_m^{(1_N 2_{N-1})}\right] \cdot \\
            & \qquad \qquad \qquad \vdots \\
            & \text{ exp}\left[-i \frac{t_2}{N} \mathcal{H}_m^{(1_2 2_2)}\right] \cdot \text{ exp}\left[-i \frac{t_1}{N} \mathcal{H}_m^{(1_2 2_1)}\right] \cdot \\
            & \text{ exp}\left[-i \frac{t_2}{N} \mathcal{H}_m^{(1_1 2_1)}\right] \cdot \text{ exp}\left[-i \frac{t_1}{N} \mathcal{H}_m^{(1_1 2_0)}\right]
        \end{split} &&
        \end{flalign}
        $U_N$ can be written in terms of an average Hamiltonian \cite{DuerNMR,HighResNMR} as
        \begin{flalign}
        \begin{split}
        & U_N = \text{ exp}\left[-i \bar{\mathcal{H}}T\right] \\
        & \text{With } T = \frac{t_1}{N} + \frac{t_2}{N} + \cdots + \frac{t_1}{N} + \frac{t_2}{N} \\
        & \text{and } \bar{\mathcal{H}} = \bar{\mathcal{H}}^0 + \bar{\mathcal{H}}^1 + \bar{\mathcal{H}}^2 + \cdots
        \end{split} &&
        \end{flalign}
        We consider only the zero order expansion and ignore higher order terms. It can be shown that the higher order terms depend inversely on $N$ and its powers and can be ignored if $N$ is sufficiently large. The average Hamiltonian is thus given by
        \begin{flalign}
        \begin{split}
        \bar{\mathcal{H}} =& \frac{1}{\frac{t_1}{N} + \frac{t_2}{N} + \cdots + \frac{t_1}{N} + \frac{t_2}{N}}\left[ \frac{t_1}{N} \mathcal{H}_m^{(1_1 2_0)} + \frac{t_2}{N} \mathcal{H}_m^{(1_1 2_1)} + \cdots + \frac{t_1}{N}\mathcal{H}_m^{(1_N 2_{N-1})} + \frac{t_2}{N}\mathcal{H}_m^{(1_N 2_N)} \right] \\
        =& \frac{1}{N(t_1 + t_2)}\left[ t_1\mathcal{H}_m^{(1_1 2_0)} + t_2\mathcal{H}_m^{(1_1 2_1)} + \cdots + t_1\mathcal{H}_m^{(1_N 2_{N-1})} + t_2\mathcal{H}_m^{(1_N 2_N)} \right] \\
        =& \frac{1}{N(t_1 + t_2)}\left[ t_1\left( \mathcal{H}_m^{(1_1 2_0)} + \mathcal{H}_m^{(1_2 2_1)} + \cdots \right) + t_2\left( \mathcal{H}_m^{(1_1 2_1)} + \mathcal{H}_m^{(1_2 2_2)} + \cdots + \right) \right] \\
        =& \frac{1}{N(t_1 + t_2)}\left[t_1\mathcal{H}_1 + t_2\mathcal{H}_2\right]
        \end{split} &&
        \end{flalign}
        $\mathcal{H}_1$ is the series sum associated with $t_1$ and $\mathcal{H}_2$ is the series sum associated with $t_2$.
        The $t_1$ series is given by
        \begin{flalign}
        \begin{split}        
        \mathcal{H}_1 =& \mathcal{H}_m^{(1_1 2_0)} + \mathcal{H}_m^{(1_2 2_1)} + \mathcal{H}_m^{(1_3 2_2)} + \cdots + \mathcal{H}_m^{(1_N 2_{N-1})} \\
        \mathcal{H}_1 =& \sum \frac{b_{ij}}{2}\left(S_i^+ S_j^+ e^{i\tau \delta_{ij}^{(1_1 2_0)}} + S_i^- S_j^- e^{-i\tau \delta_{ij}^{(1_1 2_0)}} \right) \\
        +& \sum \frac{b_{ij}}{2}\left(S_i^+ S_j^+ e^{i\tau \delta_{ij}^{(1_2 2_1)}} + S_i^- S_j^- e^{-i\tau \delta_{ij}^{(1_2 2_1)}} \right) \\
        & \qquad \qquad \qquad \vdots \\
        +& \sum \frac{b_{ij}}{2}\left(S_i^+ S_j^+ e^{i\tau \delta_{ij}^{(1_N 2_{N-1})}} + S_i^- S_j^- e^{-i\tau \delta_{ij}^{(1_N 2_{N-1})}} \right) \\
        \mathcal{H}_1 =& \sum \frac{b_{ij}}{2}S_i^+ S_j^+ \left( e^{i\tau \delta_{ij}^{(1_1 2_0)}} + e^{i\tau \delta_{ij}^{(1_2 2_1)}} + \cdots + e^{i\tau \delta_{ij}^{(1_N 2_{N-1})}} \right) \\
        +& \sum \frac{b_{ij}}{2}S_i^- S_j^- \left( e^{-i\tau \delta_{ij}^{(1_1 2_0)}} + e^{-i\tau \delta_{ij}^{(1_2 2_1)}} + \cdots + e^{-i\tau \delta_{ij}^{(1_N 2_{N-1})}} \right)
        \end{split} &&
        \label{eqn:H1}
        \end{flalign}
        The $t_2$ series is given by       
        \begin{flalign}
        \begin{split}        
        \mathcal{H}_2 =& \mathcal{H}_m^{(1_1 2_1)} + \mathcal{H}_m^{(1_2 2_2)} + \mathcal{H}_m^{(1_3 2_3)} + \cdots + \mathcal{H}_m^{(1_N 2_N)} \\
        \mathcal{H}_2 =& \sum \frac{b_{ij}}{2}\left(S_i^+ S_j^+ e^{i\tau \delta_{ij}^{(1_1 2_1)}} + S_i^- S_j^- e^{-i\tau \delta_{ij}^{(1_1 2_1)}} \right) \\
        +& \sum \frac{b_{ij}}{2}\left(S_i^+ S_j^+ e^{i\tau \delta_{ij}^{(1_2 2_2)}} + S_i^- S_j^- e^{-i\tau \delta_{ij}^{(1_2 2_2)}} \right) \\
        & \qquad \qquad \qquad \vdots \\
        +& \sum \frac{b_{ij}}{2}\left(S_i^+ S_j^+ e^{i\tau \delta_{ij}^{(1_N 2_N)}} + S_i^- S_j^- e^{-i\tau \delta_{ij}^{(1_N 2_N)}} \right) \\
        \mathcal{H}_2 =& \sum \frac{b_{ij}}{2}S_i^+ S_j^+ \left( e^{i\tau \delta_{ij}^{(1_1 2_1)}} + e^{i\tau \delta_{ij}^{(1_2 2_2)}} + \cdots + e^{i\tau \delta_{ij}^{(1_N 2_N)}} \right) \\
        +& \sum \frac{b_{ij}}{2}S_i^- S_j^- \left( e^{-i\tau \delta_{ij}^{(1_1 2_1)}} + e^{-i\tau \delta_{ij}^{(1_2 2_2)}} + \cdots + e^{-i\tau \delta_{ij}^{(1_N 2_N)}} \right)
        \end{split} &&
        \label{eqn:H2}
        \end{flalign}
        The strength of each interaction in the Hamiltonian is thus determined by the series sum in $\mathcal{H}_1$ and $\mathcal{H}_2$. Both $\mathcal{H}_1$ and $\mathcal{H}_2$ contain two series. We consider only the positive exponential series (It can be verified that the same solution is applicable to the negative exponential series). If the series sum vanishes, then that interaction is completely decoupled. The sum depends on the function $\delta_{ij}$. The following table summarizes the values taken by $\delta_{ij}$ for different interactions.
        \begin{table}[H]
\centering
\caption{$\delta_{ij}$ for Different Interactions}
\label{delta_table}
\begin{tabular}{c|c|c|}
\cline{2-3}
                                      & \textbf{Central-Peripheral}                             & \textbf{Peripheral-Peripheral}                          \\ \hline
\multicolumn{1}{|c|}{$(1_1 2_0)$}     & $\Omega_1 + \omega_1$                                   & $\omega_1 + \omega_1$                                   \\ \hline
\multicolumn{1}{|c|}{$(1_1 2_1)$}     & $\Omega_1 + \Omega_2 + \omega_1 + \omega_2$             & $\omega_1 + \omega_2 + \omega_1 + \omega_2$             \\ \hline
\multicolumn{1}{|c|}{$(1_2 2_1)$}     & $2\Omega_1 + \Omega_2 + 2\omega_1 + \omega_2$           & $2\omega_1 + \omega_2 + 2\omega_1 + \omega_2$           \\ \hline
\multicolumn{1}{|c|}{$(1_2 2_2)$}     & $2\Omega_1 + 2\Omega_2 + 2\omega_1 + 2\omega_2$         & $2\omega_1 + 2\omega_2 + 2\omega_1 + 2\omega_2$         \\ \hline
\multicolumn{1}{|c|}{$(1_3 2_2)$}     & $3\Omega_1 + 2\Omega_2 + 3\omega_1 + 2\omega_2$         & $3\omega_1 + 2\omega_2 + 3\omega_1 + 2\omega_2$         \\ \hline
\multicolumn{1}{|c|}{$(1_3 2_3)$}     & $3\Omega_1 + 3\Omega_2 + 3\omega_1 + 3\omega_2$         & $3\omega_1 + 3\omega_2 + 3\omega_1 + 3\omega_2$         \\ \hline
\multicolumn{1}{|c|}{\vdots}          & \vdots                                                  & \vdots                                                  \\ \hline
\multicolumn{1}{|c|}{$(1_N 2_{N-1})$} & $N\Omega_1 + (N-1)\Omega_2 + $ & $N\omega_1 + (N-1)\omega_2 + $ \\
\multicolumn{1}{|c|}{} & $N\omega_1 + (N-1)\omega_2$ & $N\omega_1 + (N-1)\omega_2$ \\ \hline
\multicolumn{1}{|c|}{$(1_N 2_N)$}     & $N\Omega_1 + N\Omega_2 + N\omega_1 + N\omega_2$         & $N\omega_1 + N\omega_2 + N\omega_1 + N\omega_2$         \\ \hline
\end{tabular}
\end{table}

\subsection{Peripheral-Peripheral Interactions}
    Consider peripheral-peripheral interaction first. Denoting $S_1^{PP}$ for the $t_1$ series sum and $S_2^{PP}$ for the $t_2$ series sum,
\begin{flalign}
\begin{split}
S_1^{PP} =& e^{i\tau(2\omega_1)} + e^{i\tau(4\omega_1 + 2\omega_2)} + e^{i\tau(6\omega_1 + 4\omega_2)} + \cdots + e^{i\tau(2N\omega_1 + 2(N-1)\omega_2)} \\
S_2^{PP} =& e^{i\tau(2\omega_1 + 2\omega_2)} + e^{i\tau(4\omega_1 + 4\omega_2)} + e^{i\tau(6\omega_1 + 6\omega_2)} + \cdots + e^{i\tau(2N\omega_1 + 2N\omega_2)}
\end{split} &&
\end{flalign}
$S_1^{PP}$ and $S_2^{PP}$ both are geometric series, the sum for which is given by
\begin{flalign}
\begin{split}
S_1^{PP} =& \; e^{i\tau(2\omega_1)}\left[\frac{1 - e^{iN\tau(2\omega_1 + 2\omega_2)}}{1 - e^{i\tau(2\omega_1 + 2\omega_2)}}\right] \\
=& \; e^{i\tau(2\omega_1)} \mathcal{F}_N\left(2\omega_1\tau + 2\omega_2\tau\right) \\
S_2^{PP} =& \; e^{i\tau(2\omega_1 + 2\omega_2)}\left[\frac{1 - e^{iN\tau(2\omega_1 + 2\omega_2)}}{1 - e^{i\tau(2\omega_1 + 2\omega_2)}}\right] \\
=& \; e^{i\tau(2\omega_1 + 2\omega_2)} \mathcal{F}_N\left(2\omega_1\tau + 2\omega_2\tau\right)
\end{split} &&
\end{flalign}
where $\mathcal{F}_N$ is a filter function given by
\begin{flalign}
\begin{split}
\mathcal{F}_N(x) = \frac{1 - e^{iNx}}{1 - e^{ix}}
\end{split} &&
\end{flalign}
    
    \subsection{Central-Peripheral Interactions}
    Now consider central-peripheral interaction. Denoting $S_1^{CP}$ for the $t_1$ series sum and $S_2^{CP}$ for the $t_2$ series sum, 
\begin{flalign}
\begin{split}
S_1^{CP} =& \; e^{i\tau(\Omega_1 + \omega_1)} + e^{i\tau(2\Omega_1 + \Omega_2 + 2\omega_1 + \omega_2)} + \cdots + e^{i\tau(N\Omega_1 + (N-1)\Omega_2 + N\omega_1 + (N-1)\omega_2)} \\
S_2^{CP} =& \; e^{i\tau(\Omega_1 + \Omega_2 + \omega_1 + \omega_2)} + e^{i\tau(2\Omega_1 + 2\Omega_2 + 2\omega_1 + 2\omega_2)} + \cdots + e^{i\tau(N\Omega_1 + N\Omega_2 + N\omega_1 + N\omega_2)}
\end{split} &&
\end{flalign}
Again $S_1^{CP}$ and $S_2^{CP}$ are geometric series, the sum for which is given by
\begin{flalign}
\begin{split}
S_1^{CP} =& \; e^{i\tau(\Omega_1 + \omega_1)}\left[\frac{1 - e^{iN\tau(\Omega_1 + \Omega_2 + \omega_1 + \omega_2)}}{1 - e^{i\tau(\Omega_1 + \Omega_2 + \omega_1 + \omega_2)}}\right] \\
=& \; e^{i\tau(\Omega_1 + \omega_1)} \mathcal{F}_N\left(\Omega_1\tau + \Omega_2\tau + \omega_1\tau + \omega_2\tau\right) \\
S_2^{CP} =& \; e^{i\tau(\Omega_1 + \Omega_2 + \omega_1 + \omega_2)}\left[\frac{1 - e^{iN\tau(\Omega_1 + \Omega_2 + \omega_1 + \omega_2)}}{1 - e^{i\tau(\Omega_1 + \Omega_2 + \omega_1 + \omega_2)}}\right] \\
=& \; e^{i\tau(\Omega_1 + \Omega_2 + \omega_1 + \omega_2)} \mathcal{F}_N\left(\Omega_1\tau + \Omega_2\tau + \omega_1\tau + \omega_2\tau\right)
\end{split}
\label{star_cps} &&
\end{flalign}

\subsection{Filter Function}
\begin{figure}[H]
\centering
            \def\svgwidth{0.8\linewidth}
		    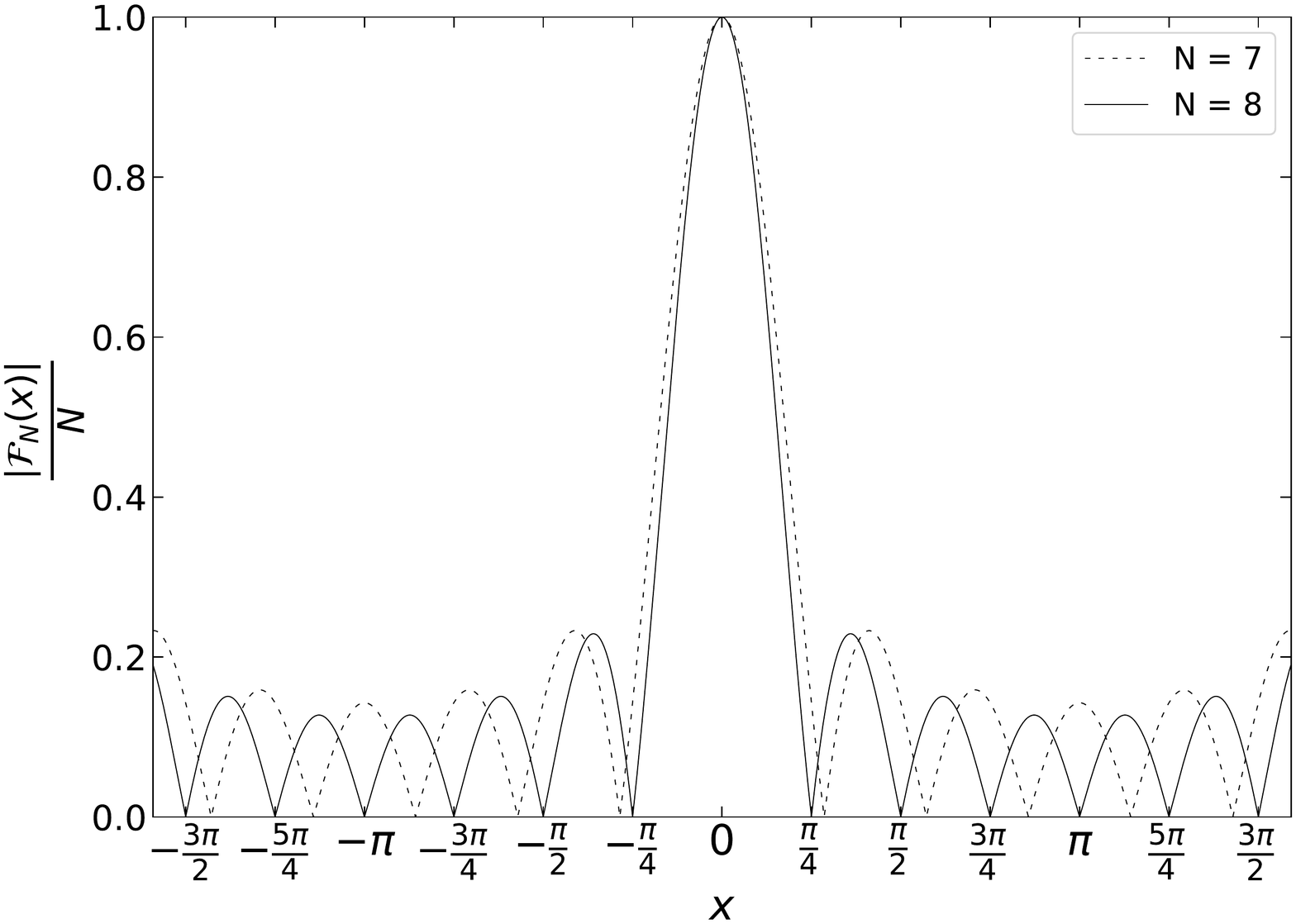
		    \caption{Filter Function for $N=7,8$}
		    \label{fig:filterfunction}
\end{figure}

The filter function $\mathcal{F}_N(x)$ can be evaluated at $x = \pi$ :
\begin{flalign}
\begin{split}
\mathcal{F}_N(\pi) =& \frac{1-e^{iN\pi}}{1-e^{i\pi}} \\
 =& \frac{1-e^{iN\pi}}{2}
\end{split}
&&
\end{flalign} 
For odd $N$, $\mathcal{F}_N(\pi)=1$ while for even $N$, $\mathcal{F}_N(\pi)=0$. In general for integer $m$, 
\begin{flalign}
\mathcal{F}_N\left((2m+1)\pi\right)= 
\begin{cases}
    1, & \text{if N is odd} \\
    0, & \text{if N is even}
\end{cases}
&&
\end{flalign}
Figure \ref{fig:filterfunction} shows the filter function plotted for $N=7,8$.

		\subsection{Decoupling Conditions}
The filter function peaks at $x=2n\pi$ taking the value $N$ and takes the value of $0$ at $x=(2n+1)\pi$ for even $N$. To decouple the peripheral-peripheral interactions, the argument of the filter function thus needs to be an odd integer multiple of $\pi$.
\begin{itemize}
\item \textbf{Condition 1 : } $ (2\omega_1 + 2\omega_2)\tau = (2n + 1)\pi $
\end{itemize}
Similarly, to retain the central-peripheral coupling, the argument of the filter function should be an even integer multiple of $\pi$.
\begin{itemize}
        \item \textbf{Condition 2 : } $(\Omega_1 + \Omega_2 + \omega_1 + \omega_2)\tau = 2m\pi$
        \end{itemize}
        Putting Condition 2 back in equation \ref{star_cps},
\begin{flalign}
\begin{split}
\label{eqn:extra_cond}
    S_1^{CP} =& \; N e^{i\tau(\Omega_1 + \omega_1)} \\
    S_2^{CP} =& \; N
\end{split} &&
\end{flalign}
One can impose an additional constraint to make $S_1^{CP} = N$ which implies that $(\Omega_1 + \omega_1)\tau = 2m_1\pi$. This effectively reduces the second condition to $(\Omega_2 + \omega_2)\tau = 2m_2\pi$.

\noindent
\textbf{Conditions for $L=2$ : }
\begin{flalign}
\begin{split}
    &(\Omega_1 + \omega_1)\tau = 2m_1\pi \\
    &(\Omega_2 + \omega_2)\tau = 2m_2\pi \\
    &(2\omega_1 + 2\omega_2)\tau = (2n + 1)\pi
\end{split} &&
\end{flalign}
    \textbf{General Conditions : } \\
    In general for $L$ stages it can be shown that there are $L+1$ conditions. We only provide the final conditions which can be worked out in a fashion similar as above (See Appendix \ref{appendix:gdc}).
\begin{flalign}
\begin{split}
    &2\sum_{i=1}^L \omega_i\tau = (2l + 1)\pi \\
    &(\Omega_i + \omega_i)\tau = 2m_i\pi \;\;\;\;\text{with i running from 1 to L}
\end{split}
\label{star_generalcond} &&
\end{flalign}

\subsection{Average Hamiltonian}
Using these conditions, we compute $\mathcal{H}_1$ and $\mathcal{H}_2$ from Equations \ref{eqn:H1} and \ref{eqn:H2}. 
    \begin{flalign}
    \begin{split}
    \mathcal{H}_1 =& N\sum_{j=2}^n \frac{b_{1j}}{2}\left(S_1^+ S_j^+ + S_1^- S_j^- \right) = N\mathcal{H}_0 \\
    \mathcal{H}_2 =& N\sum_{j=2}^n \frac{b_{1j}}{2}\left(S_1^+ S_j^+ + S_1^- S_j^- \right) = N\mathcal{H}_0
    \end{split} &&
    \end{flalign}
 The average Hamiltonian reduces to
\begin{flalign}
\begin{split}
\bar{\mathcal{H}} =& \frac{1}{N(t_1 + t_2)}\left[ Nt_1\mathcal{H}_0 + Nt_2\mathcal{H}_0 \right] \\
=& \; \mathcal{H}_0
\end{split} &&
\end{flalign}
where
\begin{flalign}
\begin{split}
\label{avgham}
    \mathcal{H}_0 =& \sum_{j=2}^n \frac{b_{1j}}{2} \left( S_1^+S_j^+ + S_1^-S_j^- \right) \\
    =& \sum_{j=2}^n b_{1j} \left(S_1^xS_j^x - S_1^yS_j^y\right)
\end{split} &&
\end{flalign}
This is the DQ Hamiltonian with only the radial interactions present. Thus the average Hamiltonian over the entire time period completely decouples the peripheral-peripheral interactions.

\section{Simulations}
\begin{figure}[H]
\centering
  \def\svgwidth{0.8\linewidth}
		    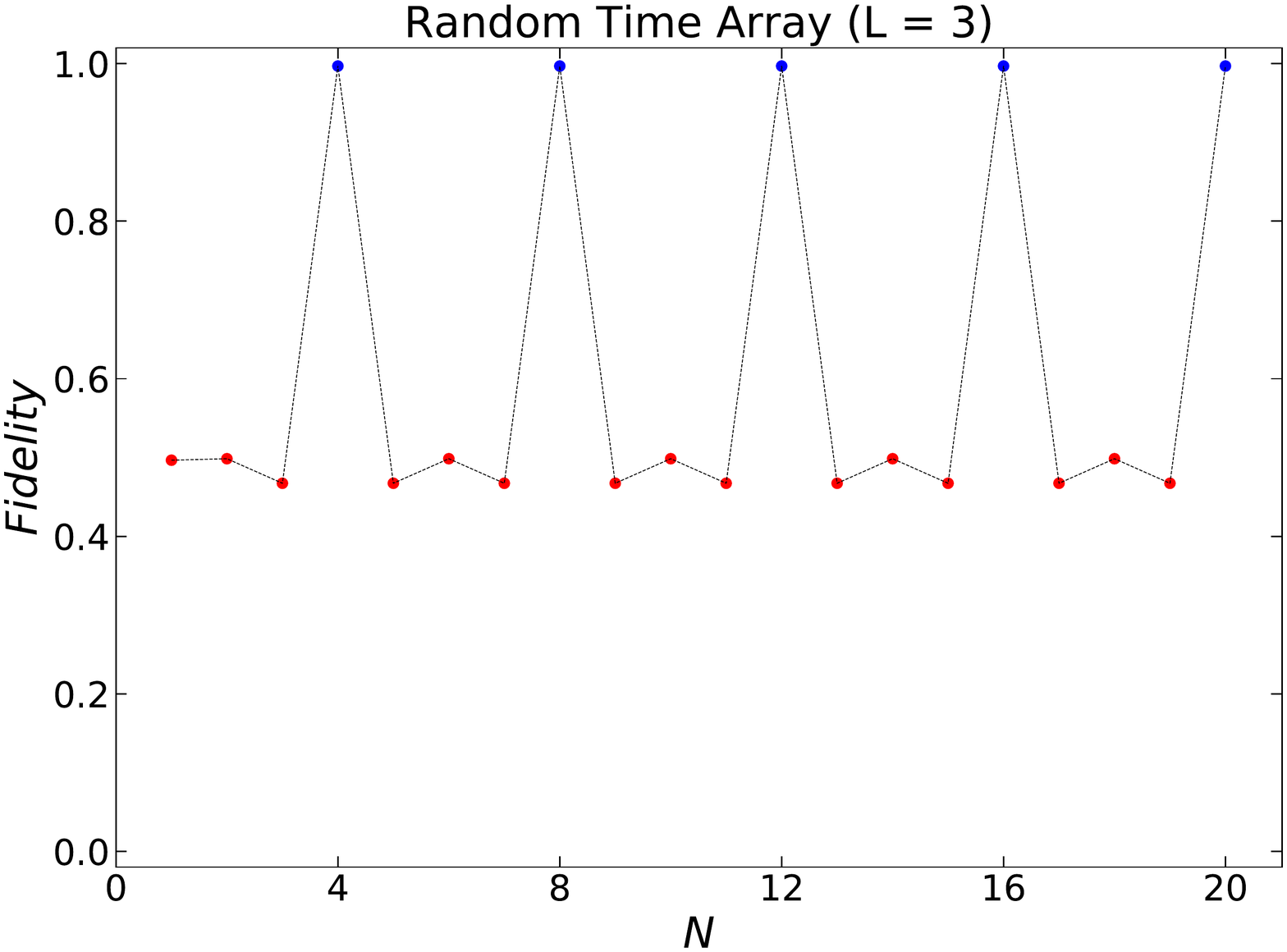
		    \caption{Fidelity Profile for a random time array $[t_1,t_2,t_3]$ where $t_i\in(0,0.1]$.}
		    \label{fig:SFP_RT}
\end{figure}

The general conditions \ref{star_generalcond} derived in previous section place no conditions on $\tau$ or $\omega$ alone but the product of them. For simplicity, we assume $\tau=1$ and work only with the frequencies. If a different value is taken for $\tau$, the new required frequencies can be easily worked out. Conditions for perfect decoupling are
\begin{flalign}
\begin{split}
\label{sim_conditions}
2\sum_{i=1}^L \omega_i =& \; (2l+1)\pi \\
\Omega_i + \omega_i =& \; 2m_i\pi
\end{split} &&
\end{flalign}
We choose the following solution set for equations \ref{sim_conditions}
\begin{flalign}
\begin{split}
\label{sim_params}
\omega_i =& \; \omega_j = \omega = \left(\frac{2L+1}{2L}\right)\pi \\
\Omega_i =& \; \Omega_j = \Omega = -\omega
\end{split} &&
\end{flalign}
The fact that these parameters satisfy the conditions can be easily verified. We simulate for $L=3$, for which the parameters are
\begin{flalign}
\begin{split}
\omega_i =& \; \omega_j = \frac{7}{6}\pi \\
\Omega_i =& \; \Omega_j = -\frac{7}{6}\pi
\end{split} &&
\end{flalign}
The filtered scheme places no restrictions on the times $t_1$, $t_2$,.., $t_L$. We denote this as the time array $[t_1,t_2,..,t_L]$ (including the $N$ term in the denominator, namely $\frac{t_i}{N}$). Simulations \cite{QuTiP} are done for two cases - A constant time array and a random time array drawn from a uniform distribution. We simulate the system for $N=20$. Based on the filter function characteristics, we expect peak fidelity for even cycles. However, we observe peak fidelity only for every $4^{th}$ cycle.
\begin{figure}[H]
    \centering
  \def\svgwidth{0.8\linewidth}
		    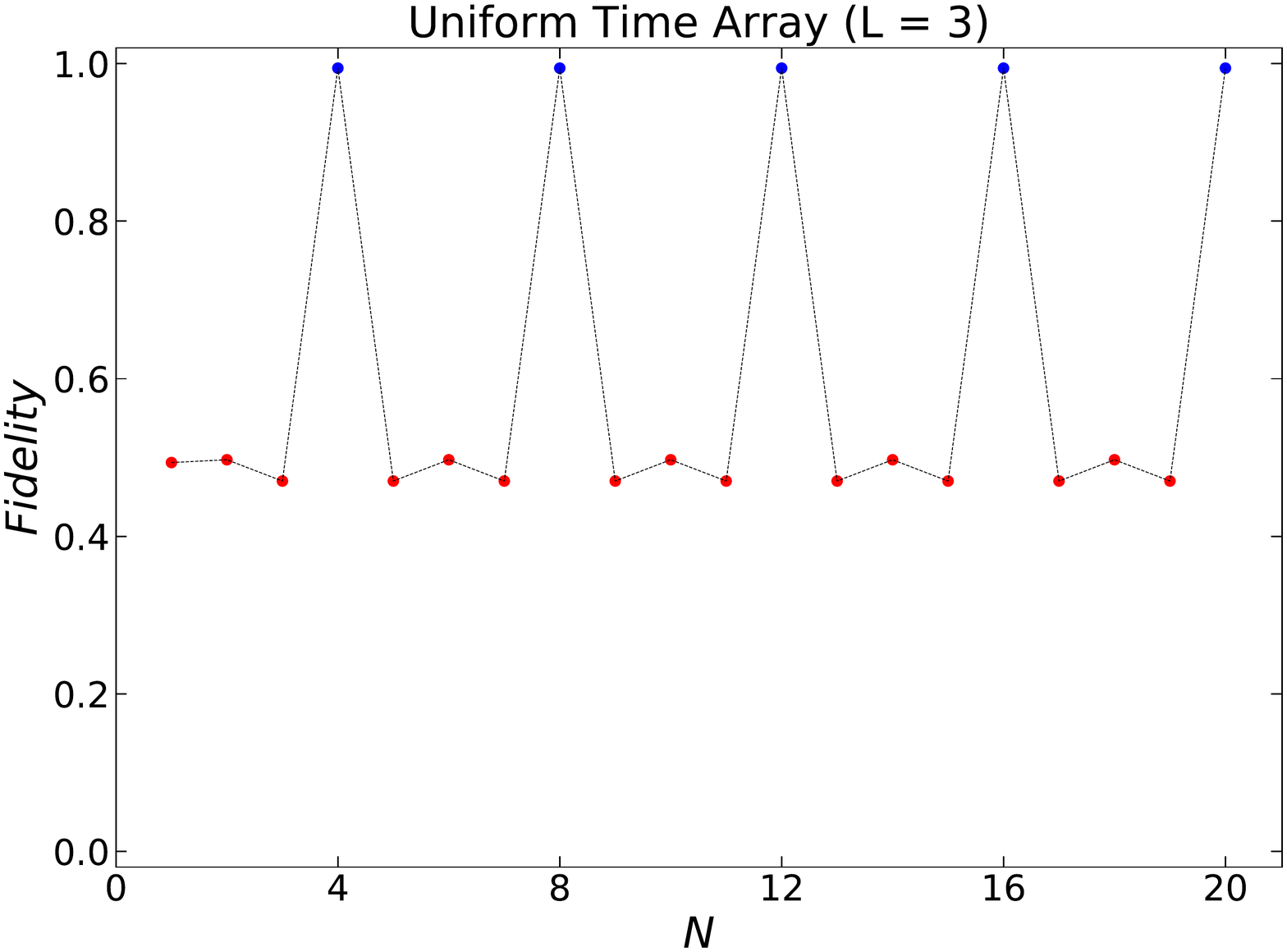
		    \caption{Fidelity Profile for a uniform time array $[t,t,t]$ where $t=0.05$.}
		    \label{fig:SFP_UT}
\end{figure}
\noindent
The random time array is taken from a uniform distribution bounded by $0$ and $0.1$. Figure \ref{fig:SFP_RT} shows the results.
\begin{figure}[H]
\centering
  \def\svgwidth{0.8\linewidth}
		    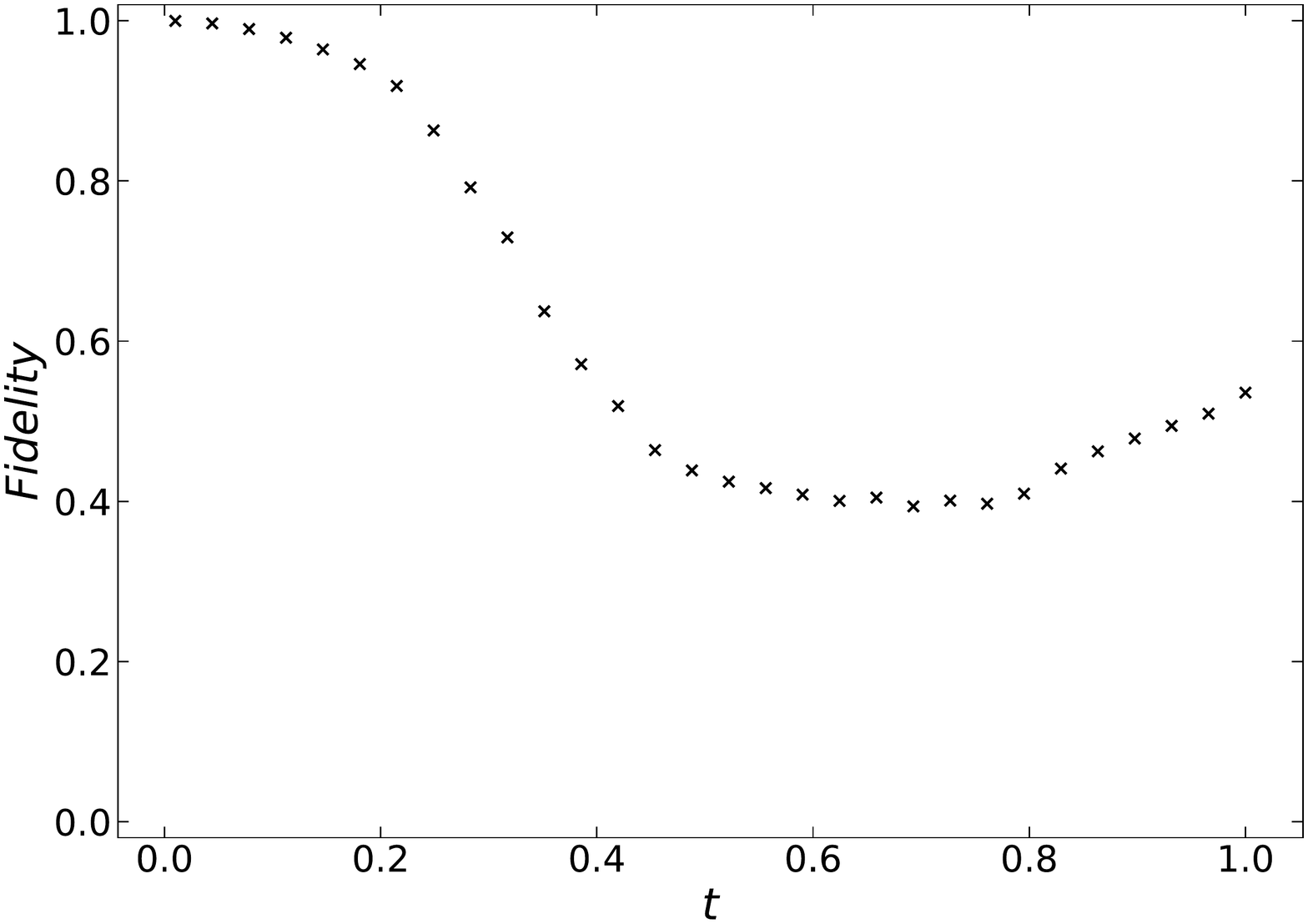
		    \caption{Fidelity vs $t$ for a constant time array $[t,t,t]$}
		    \label{fig:TimePerformance}
\end{figure}
\noindent
Simulation results show that the filtered scheme is not dependent on specific values of time taken in the time array and hence is very robust. However, it is ineffective if $t$ is too large. To study the dependence of the scheme on time, we perform simulations for different values of t in a constant time array with $L=3$. We consider fidelity at $N=8$ as we assume that by that cycle, the filtered scheme should achieve its purpose. Figure \ref{fig:TimePerformance} shows the performance of the scheme as a function of $t$ in constant time array $[t,t,t]$. The scheme starts losing efficiency for values of $t$ greater than around $0.1$. This is justified since only zero order terms in the average Hamiltonian were considered. As $t$ increases, higher order terms contribute more. This was done for $L=3$. For higher $L$, the effects are more pronounced.
	\section{Conclusion}
	\label{sec:star_concl}
    The filtered engineering technique can be used to create a star topology. Since this method effectively decouples homo-nuclear interactions, homo-nuclear decoupling pulse sequences \cite{DuerNMR} can be applied to get the same result. The filtered technique can also be used for selective decoupling. If there are 3 kinds of distinguishable spin species in the model, then such a technique can be used to selectively decouple specific interactions. For example, if we denote the three species as A, B, C and we want to decouple A-A but retain B-B and C-C, a filtered technique can be used to do this. 

    In terms of the efficiency, the scheme is very efficient for small time-scales but the fidelity starts dropping for larger times. The scheme is also quite robust in terms of the values of time in the time array. As long as the total time is not too large, the scheme will create a star topology with very high fidelity.	
	
	\chapter{Quantum Information Transport}
	\section{Chain}
	A spin chain is a perfect short-range communication device for quantum computers. We consider here a simple chain with uniform couplings and magnetic fields only on the ends of the chain. This creates a resonance effect in the sense that the free ends of the spin communicate with each other. We show that information transport can be achieved in this simple architecture.
\begin{figure}[H]
\centering
\begin{tikzpicture}
\begin{scope}[every node/.style={circle,thick,draw}]
    \node (1) at (0,0) {1};
    \node (2) at (2,0) {2};
    \node (3) at (4,0) {3};
\end{scope}

\begin{scope}[>={Stealth[black]},
              every node/.style={fill=white,circle},
              every edge/.style={draw=black,thick}]
    \path [-] (1) edge (2);
    \path [-] (2) edge (3);
\end{scope}
\end{tikzpicture}
\caption{3-Spin Chain}
\label{fig:img_chain}
\end{figure}
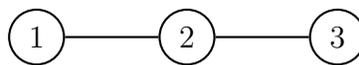
    \subsection{System}
    Consider a spin chain governed by the XY Hamiltonian.
    \begin{flalign}
  \mathcal{H} = h \left(S_1^z+S_n^z\right) + J \sum_{i} \left(S_i^xS_{i+1}^x + S_i^yS_{i+1}^y\right) &&
  \end{flalign}
    Since the XY Hamiltonian commutes with the total spin operator,
    \begin{flalign}
     \left[\mathcal{H}, \sum_i S_i^z\right]=0&&
    \end{flalign}
    the total spin number is conserved. The state $|000..0\rangle$ is a stationary state. It suffices to show that the state $|100..0\rangle$ evolves to the state $|000..1\rangle$ under natural evolution of the system as it implies that any arbitrary state $\alpha|0\rangle + \beta|1\rangle$ is transported.
    
    For simplicity, consider a 3-spin chain. We consider the dynamics in the 1-excitation space only. The basis for this subspace is $|001\rangle$, $|010\rangle$ and $|100\rangle$. Hamiltonian in this basis is
    \begin{flalign}
    \mathcal{H}
    =\frac{1}{2}
    \begin{pmatrix}
    0 & J & 0 \\
    J & h & J \\
    0 & J & 0
    \end{pmatrix}    
     &&
    \end{flalign}
    To obtain an optimum time when transport takes place, we solve the eigenvalue problem for that time. The eigenvalues and eigenvectors are given by
    \begin{flalign}
\begin{split}
    \lambda_1 =& \; 0 \\
    \lambda_2 =& \; \frac{1}{2}\left(h-\sqrt{h^2+2J^2}\right) \\
    \lambda_2 =& \; \frac{1}{2}\left(h+\sqrt{h^2+2J^2}\right)
\end{split} &&
\end{flalign}
and the corresponding eigenvectors are
\begin{flalign}
    v_1
    ={
    \begin{pmatrix}
    -1 \\
    0 \\
    1 
    \end{pmatrix}}
    \;\;v_2
    ={
    \begin{pmatrix}
    1 \\
    \frac{h-\sqrt{h^2+2J^2}}{J} \\
    1    
    \end{pmatrix}}
    \;\;v_3
    ={
    \begin{pmatrix}
    1 \\
    \frac{h+\sqrt{h^2+2J^2}}{J} \\
    1    
    \end{pmatrix}} &&
\end{flalign}
System starts in state $|100\rangle = \begin{pmatrix}
    0 \\
    0 \\
    1 
    \end{pmatrix}$ \\
and it evolves to the state $|001\rangle = \begin{pmatrix}
    1 \\
    0 \\
    0 
    \end{pmatrix}$ after some time $\tau$.\\
    We have
    \begin{flalign}
    e^{-i\mathcal{H}\tau}\begin{pmatrix}
    0 \\
    0 \\
    1 
    \end{pmatrix}=
    \begin{pmatrix}
    1 \\
    0 \\
    0 
    \end{pmatrix}
     &&
    \end{flalign}
Writing the initial state in terms of the eigenvectors of the Hamiltonian, we have
\begin{flalign}
    \begin{pmatrix}
    0 \\
    0 \\
    1 
    \end{pmatrix} =
    c_1v_1+c_2v_2+c_3v_3
     &&
    \end{flalign}
    with
    \begin{flalign}
    \begin{split}
    c_1 =& \; \frac{1}{2}\\
    c_2 =& \; \frac{h+\sqrt{h^2+2J^2}}{4\sqrt{h^2+2J^2}}\\
    c_3 =& \; \frac{-h+\sqrt{h^2+2J^2}}{4\sqrt{h^2+2J^2}}
    \end{split}
     &&
    \end{flalign}
    The final state is thus
    \begin{flalign}
    \begin{pmatrix}
    1 \\
    0 \\
    0 
    \end{pmatrix} =
    c_1e^{-i\lambda_1\tau}v_1+c_2e^{-i\lambda_2\tau}v_2+c_3e^{-i\lambda_3\tau}v_3
     &&
    \end{flalign}
    
	\subsection{Transport Conditions}
	Solving for $\tau$, we get the following conditions
	\begin{flalign}
    \begin{split}
    e^{-i\lambda_2\tau} = -1 \\
    e^{-i\lambda_3\tau} = -1
    \end{split}
     &&
    \end{flalign}
    which implies
    \begin{flalign}
    \begin{split}
    \lambda_2\tau = \left(2n_1+1\right)\pi \\
    \lambda_3\tau = \left(2n_2+1\right)\pi
    \end{split}
     &&
    \end{flalign}
    The time required is constrained by two integers and is not generally solvable. However, we can plot $cos(\lambda_2\tau)$ and $cos(\lambda_3\tau)$ and check when both are simultaneously $-1$. For example, consider $J = 2\pi*10$ and $h = 2\pi * 100$.
    \begin{figure}[H]
    \centering
            \def\svgwidth{0.8\linewidth}
		    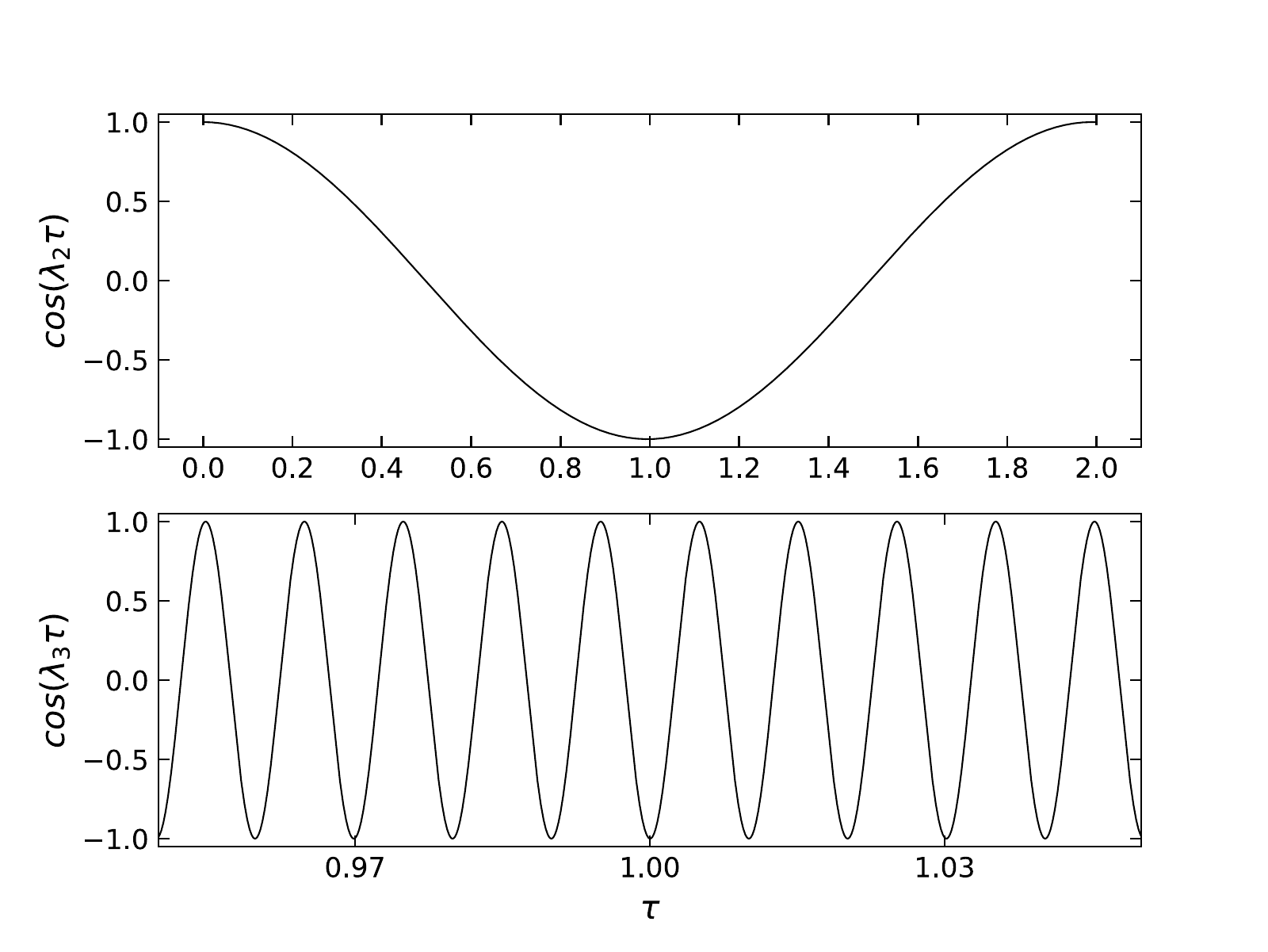
		    \caption{Chain - $cos(\lambda_2\tau)$ and $cos(\lambda_3\tau)$ vs $\tau$.}
		    \label{fig:chain_opt_time}
\end{figure}
\noindent
$cos(\lambda_2\tau)$ and $cos(\lambda_3\tau)$ are simultaneously -1 at $\tau=1$. $cos(\lambda_3\tau)$ is a rapidly oscillating function of $\tau$ since the argument is very small in magnitude. In Figure \ref{fig:chain_opt_time}, we have shown its relevant profile around $\tau=1$. In general, the time required for transport is dependent on the chain length. Similar solutions can be worked out for higher lengths. We discuss modularity later where one could attach two such chains and achieve their combined purpose. However, time dependent magnetic fields are required in such a scenario.
\subsection{Simulations}
We simulate the system with same parameters. We expect that at $\tau=1$, we get optimal transport. We consider state transport for two input states :
\begin{enumerate}
\item $|1\rangle$
\item $|+\rangle=\frac{1}{\sqrt{2}}\left(|0\rangle+|1\rangle\right)$
\end{enumerate}
\begin{figure}[H]
\centering
            \def\svgwidth{0.8\linewidth}
		    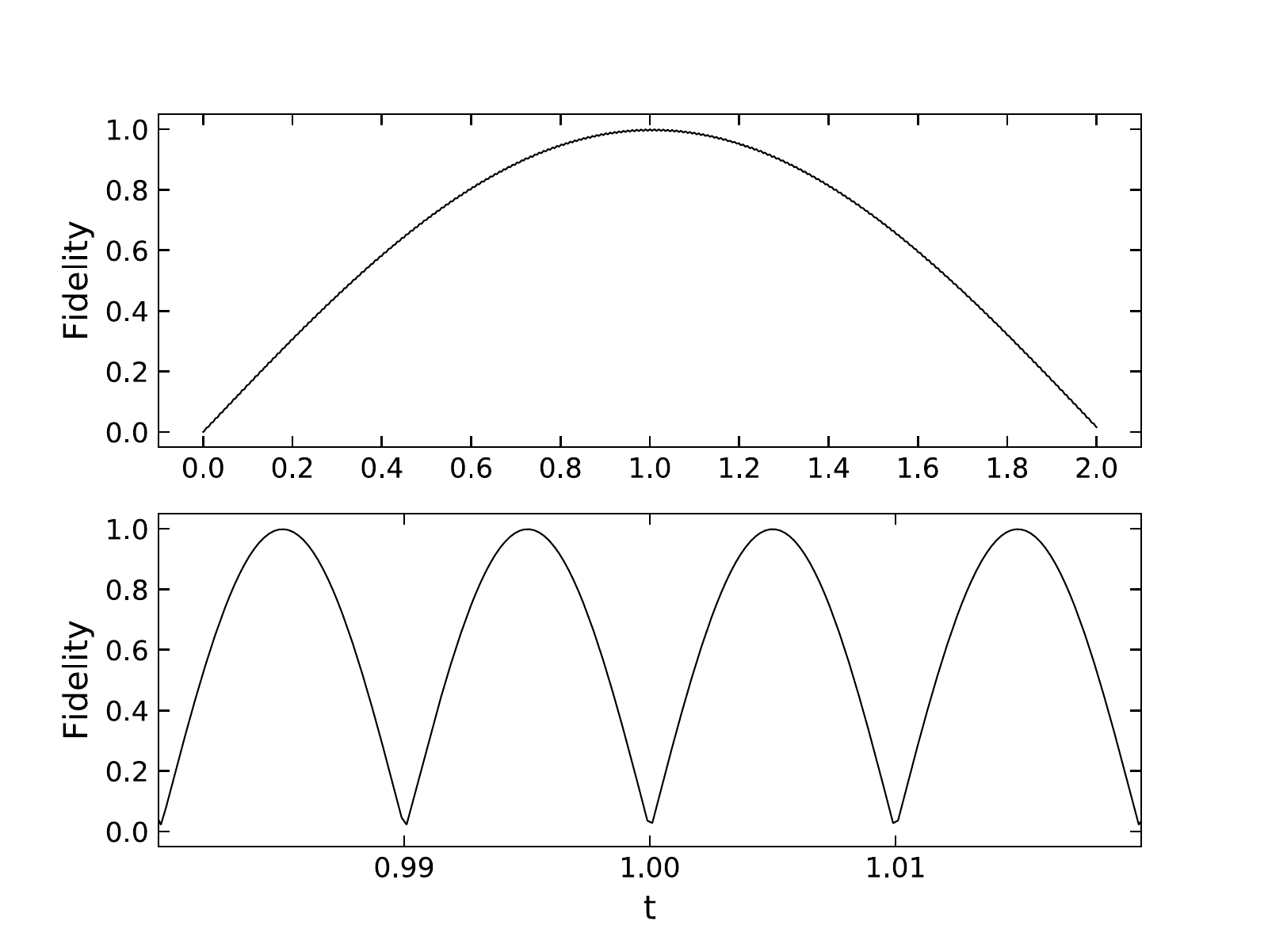
		    \caption{Chain - Fidelity vs $t$. Top : Input State $|1\rangle,\;\;$ Bottom : Input State : $|+\rangle\;\;$.}
		    \label{fig:chain3}
\end{figure}
For superposition transport, the fidelity is a rapidly oscillating function of time. We have shown the relevant region around $t=1$ in Figure \ref{fig:chain3}. A possible reason for this could be the introduction of a relative phase during evolution. Peak fidelity for superposition transport occurs at $t=0.995$ and $t=1.005$ and not at $t=1$. We thus consider optimal time to be $\tau=1.005$ and not $\tau=1$. To verify that any arbitrary state is transferred with peak fidelity at this time, we find the mean of the fidelity obtained for a large number of states on the Bloch sphere. A general state defined on the Bloch sphere is \cite{NielsenChuang}
\begin{flalign}
    |\psi\rangle = cos\left(\frac{\theta}{2}\right)|0\rangle+e^{i\phi}sin\left(\frac{\theta}{2}\right)|1\rangle   
     &&
    \end{flalign}
    where $0\leq\theta\leq\pi$ and $0\leq\phi<2\pi$. We consider a meshgrid of 100 $\theta$ values and 100 $\phi$ values lying uniformly in their respective domains. This gives $10000$ unique states on the Bloch sphere. We got a mean fidelity of $0.9981$ with a standard deviation of $0.0018$. Maximum fidelity obtained was $1$ while the minimum was $0.9951$. The optimal time is quite robust for any arbitrary state. We further study robustness of the model under perturbations in Hamiltonian parameters. 

	\subsection{Robustness}
	We study the robustness of the scheme under perturbations in the Hamiltonian parameters. We consider 3 parameters for perturbations : $h_1$, $h_2$, $J_{12}$. As before, we use two input states for the analyses : $|1\rangle$ and $|+\rangle$.
	\begin{figure}[H]
    \centering
  \def\svgwidth{0.8\linewidth}
		    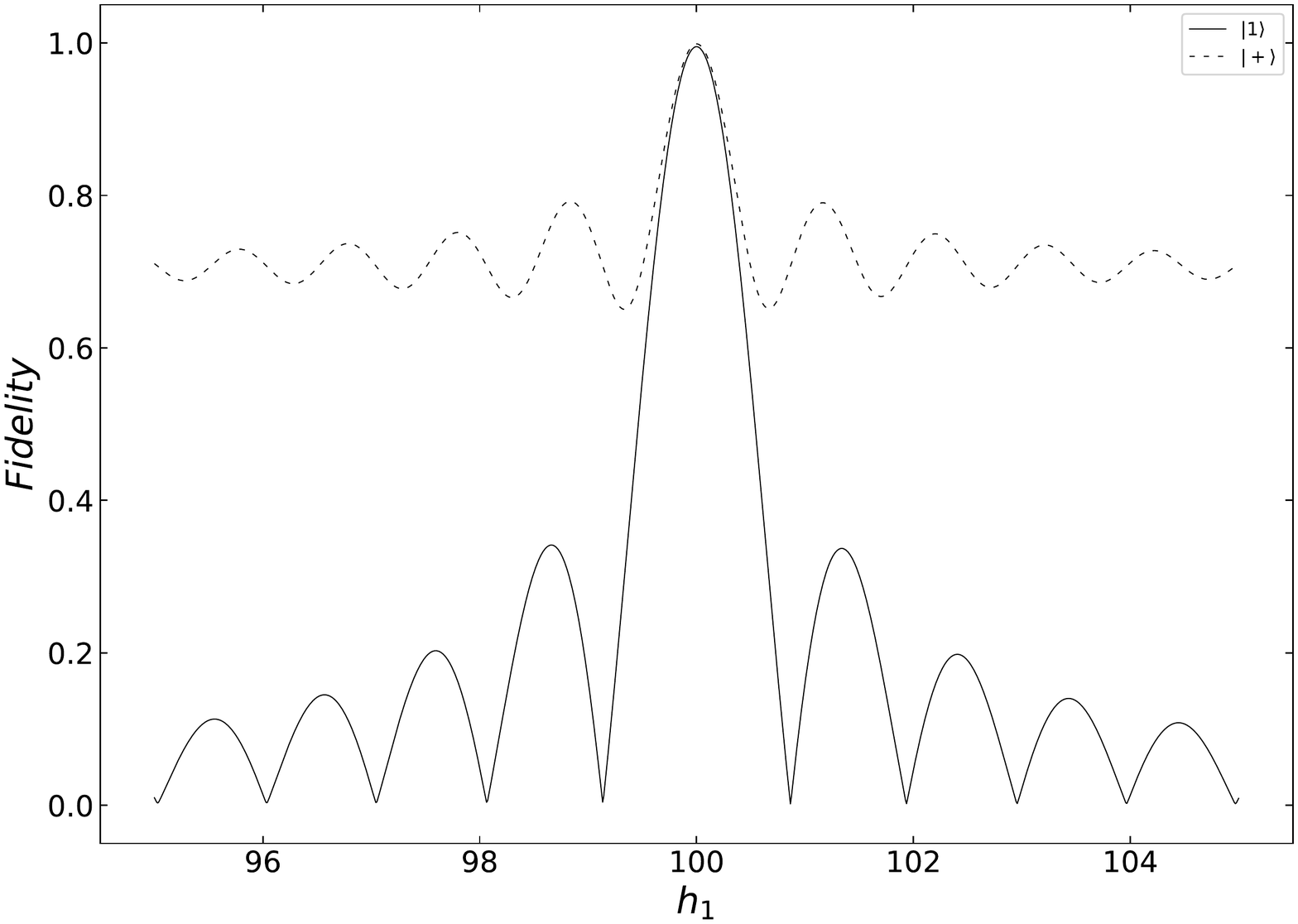
		    \caption{Chain Robustness analysis - Fidelity vs $h_1$}
		    \label{fig:chain_rob_h1}
\end{figure}
\noindent
There is a strong drop off in fidelity for $h_1$ perturbation. This is expected since it is responsible to establish the resonance with the last spin.
\begin{figure}[H]
    \centering
  \def\svgwidth{0.8\linewidth}
		    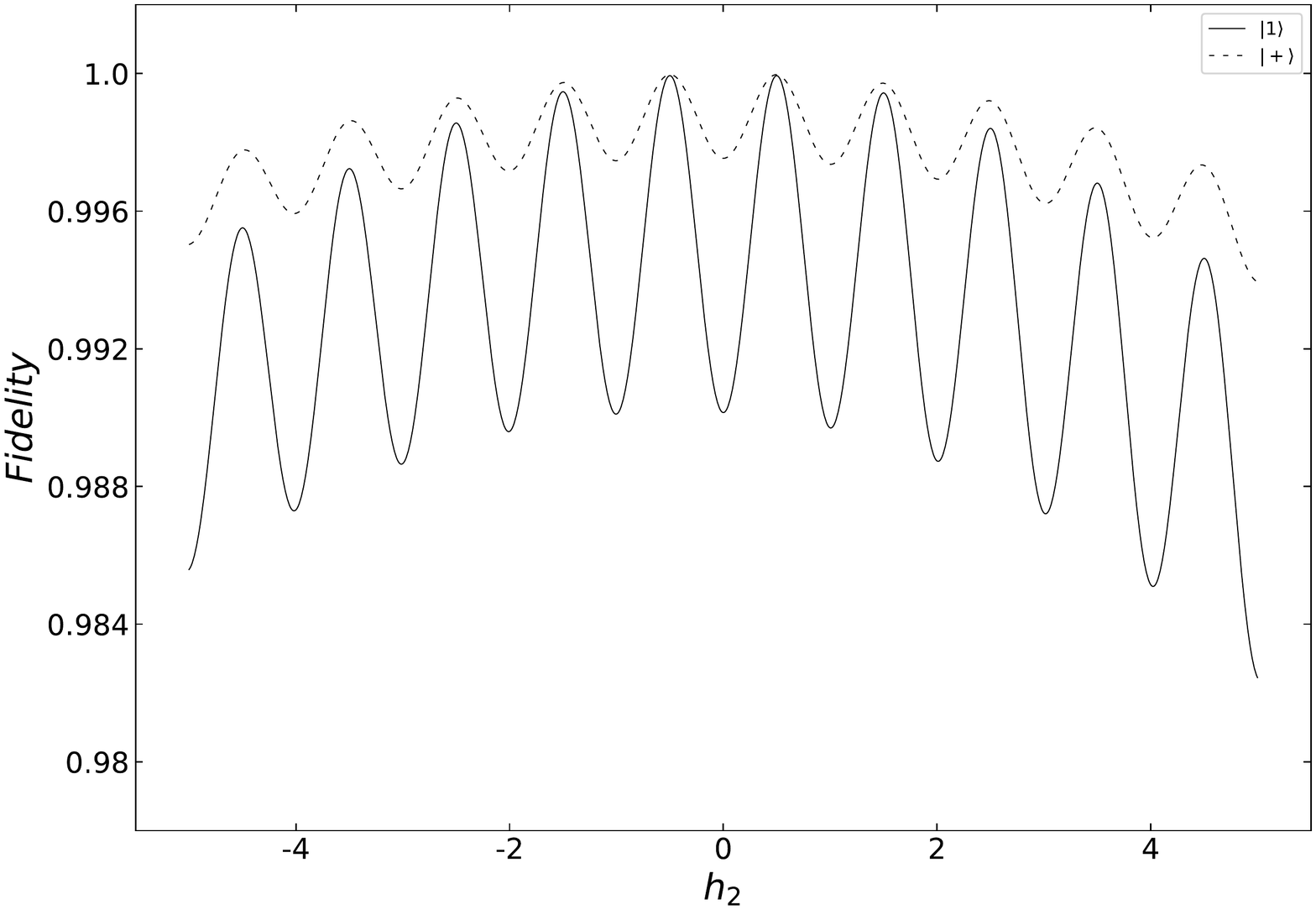
		    \caption{Chain Robustness analysis - Fidelity vs $h_2\;\;$}
		    \label{fig:chain_rob_h2}
\end{figure}
\begin{figure}[H]
    \centering
  \def\svgwidth{0.8\linewidth}
		    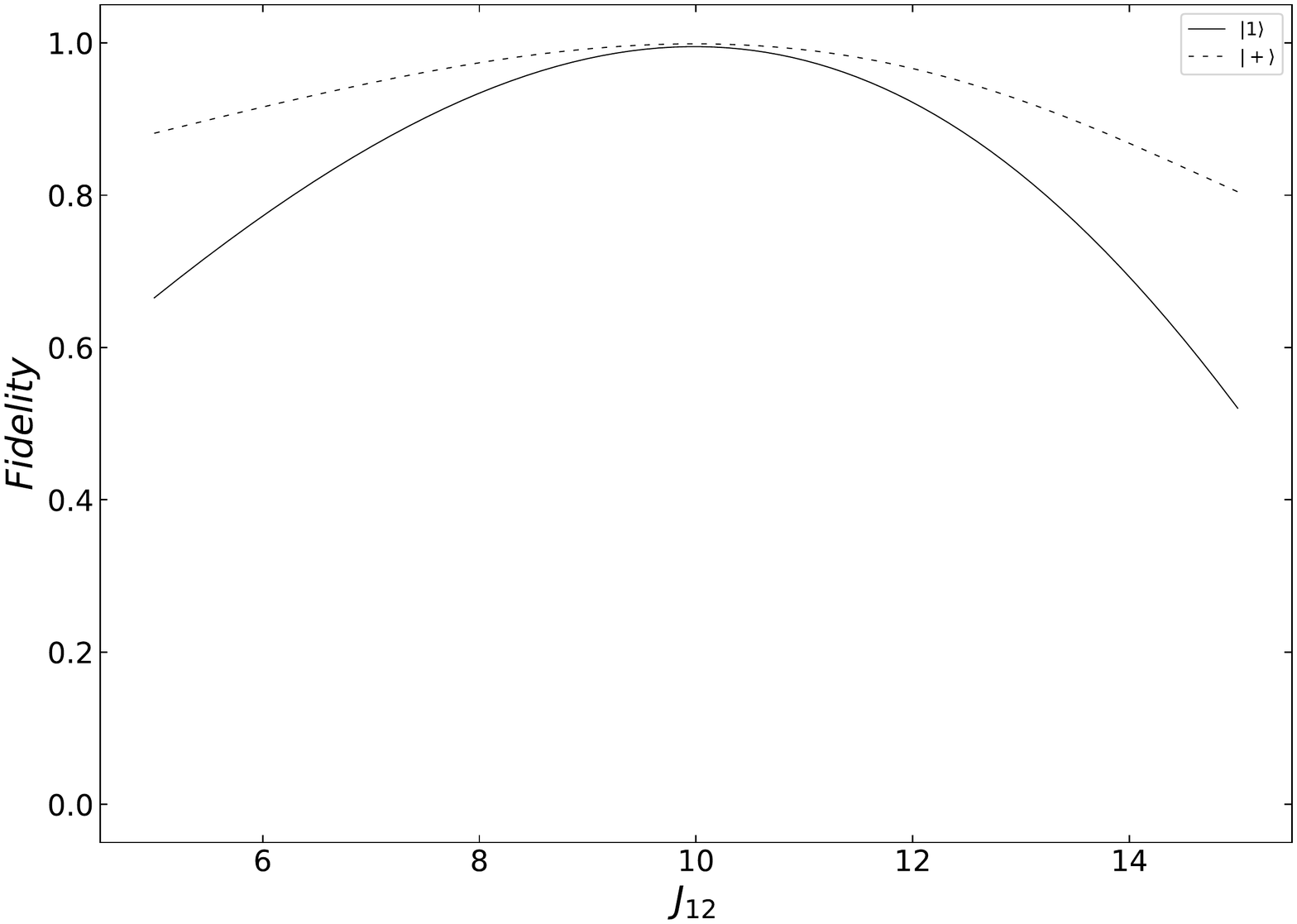
		    \caption{Chain Robustness analysis - Fidelity vs $J_{12}$}
		    \label{fig:chain_rob_J1}
\end{figure}
\noindent
Compared to $h_1$, perturbations in $h_2$ result in negligible drop in fidelity as seen in Figure \ref{fig:chain_rob_h2} (The fidelity scale begins at $0.98$). The central spins are thus very much robust. However, we observe that peak fidelity does not occur at $h_2=0$ but at the side-lobes around it. The scheme can be thus further optimized by setting $h_2$ to these values. Figure \ref{fig:chain_rob_J1} shows the fidelity profile against perturbations to the coupling strength. The coupling is much more robust compared to $h_1$.
	\section{4-spin Router}
	The resonance effect in a chain can be extended to a routing mechanism. Instead of a single output node, we consider an additional output node with a magnetic field different to that of the original output node. In such a 4-spin system with uniform couplings we show that with minimal control, one can route information from the central node towards one of the two output nodes. This can be extended to a general star topology to achieve a switch mechanism \cite{SpinStarSwitch}.
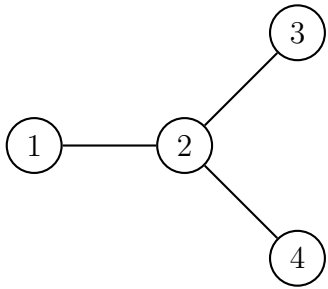
\begin{figure}[H]
\centering
\begin{tikzpicture}
\begin{scope}[every node/.style={circle,thick,draw}]
    \node (1) at (0,0) {1};
    \node (2) at (2,0) {2};
    \node (3) at (3.5,1.5) {3};
    \node (4) at (3.5,-1.5) {4};
\end{scope}

\begin{scope}[>={Stealth[black]},
              every node/.style={fill=white,circle},
              every edge/.style={draw=black,thick}]
    \path [-] (1) edge (2);
    \path [-] (2) edge (3);
    \path [-] (2) edge (4);
\end{scope}
\end{tikzpicture}
\label{fig:img_router}
\caption{4-Spin Router}
\end{figure}
	\subsection{System}
	The system is a 4-spin network in a star topology. One of the peripheral nodes is the input node while the other two are the output nodes. The objective is to route information from the input node to one of the output nodes based on some control. The Hamiltonian is given by
	\begin{flalign}
  \mathcal{H} = \sum_ih_iS_i^z + J \sum_{i,j} \left(S_i^xS_j^x + S_i^yS_j^y\right) &&
  \end{flalign}
	Analogous to the solution to the spin chain, we consider uniform couplings and zero magnetic field on the central spin. For the output spins, we set their magnetic fields to be $+h$ and $-h$. We show that setting the input spin field to $+h$ results in resonant transfer to the first output spin while the second output spin is off-resonance. Switching the the input field to $-h$ results in the opposite with the first output spin remaining off-resonant. Analytical solution for this scheme is difficult to work out. Instead we show simulation results which confirm that this routing mechanism works. In the next section, we consider a 5-spin router for which analytical solution is worked out.
	\subsection{Simulations}
	\begin{figure}[H]
\centering
            \def\svgwidth{0.8\linewidth}
		    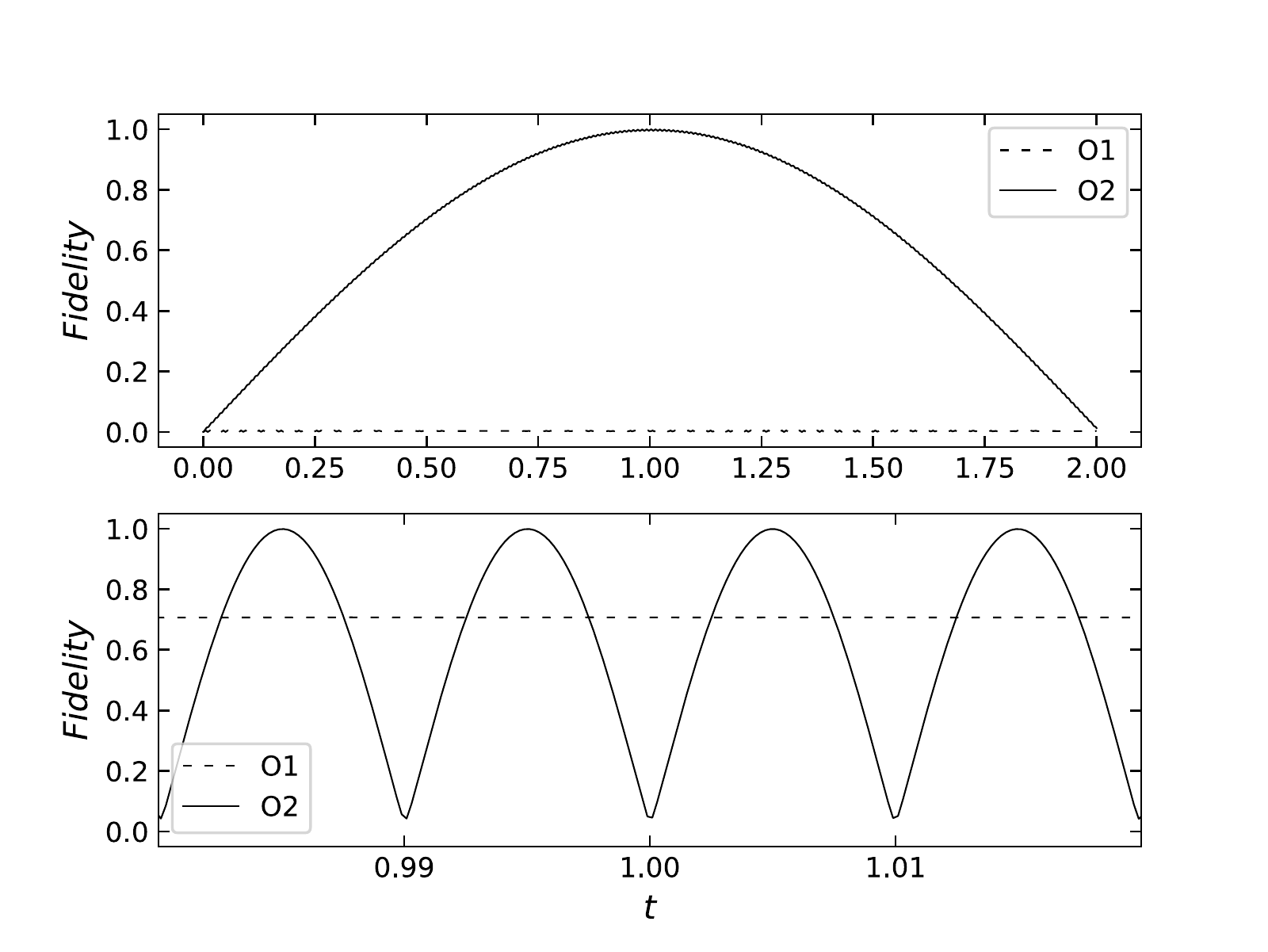
		    \caption{4-Spin Router - Fidelity vs $t$ for normal case}
		    \label{fig:router4}
\end{figure}
	We use similar parameters used in the chain problem namely  $h=2\pi*100$ and $J=2\pi*10$. As before, we consider the same two input states for transport : $|1\rangle$ and $|+\rangle$.
\begin{figure}[H]
\centering
            \def\svgwidth{0.8\linewidth}
		    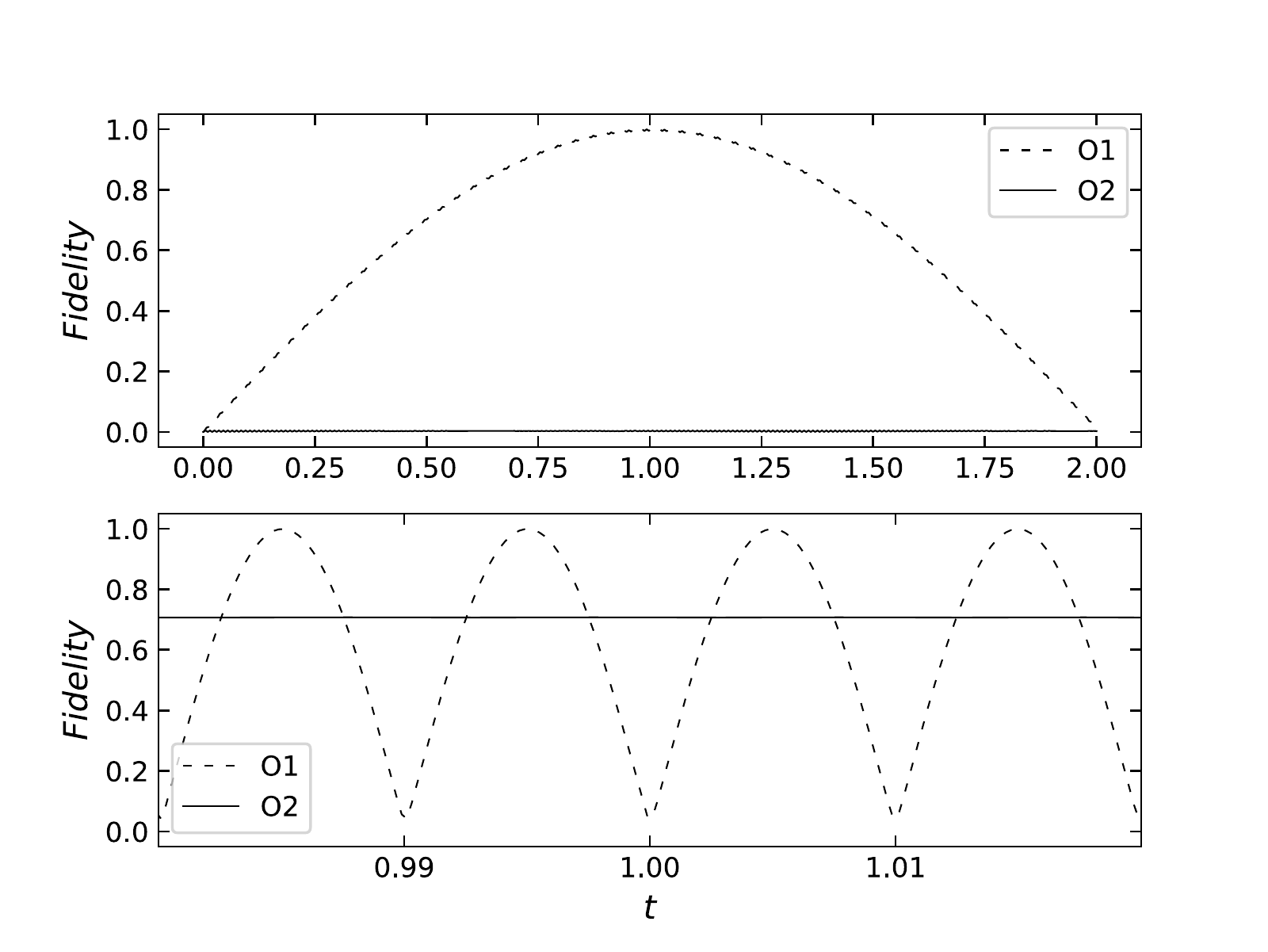
		    \caption{4-Spin Router - Fidelity vs $t$ for switched case}
		    \label{fig:router4_switched}
\end{figure}
\newpage
	
	\section{5-spin Router}
	Markuchov et. al. \cite{SpinTransistor} showed conditional state transfer in a Heisenberg spin chain where they use the central two spins as a gate that controls the flow of information. We extend their idea here and show routing mechanism in a 5 spin system.
\begin{figure}[H]
\centering
\begin{tikzpicture}
\begin{scope}[every node/.style={circle,thick,draw}]
    \node (1) at (0,0) {1};
    \node (2) at (2,0) {2};
    \node (3) at (4,0) {3};
    \node (4) at (5.5,1.5) {4};
    \node (5) at (5.5,-1.5) {5};
\end{scope}

\begin{scope}[>={Stealth[black]},
              every node/.style={fill=white,circle},
              every edge/.style={draw=black,thick}]
    \path [-] (1) edge (2);
    \path [-] (2) edge (3);
    \path [-] (3) edge (4);
    \path [-] (3) edge (5);
\end{scope}
\end{tikzpicture}
\label{fig:img_router5}
\caption{5-Spin Router}
\end{figure}
	\subsection{System}
The system is governed by the XY Hamiltonian given by
    \begin{flalign}
    \mathcal{H} = \sum_ih_iS_i^z - \sum_{l, m}J_{lm}\left(S_l^xS_m^x + S_l^yS_m^y\right) &&
    \end{flalign}
    where $(l, m) \in \{(1, 2), (2, 3), (3, 4), (3, 5) \}$
    \medskip
    
    As before, since the total spin for the XY Hamiltonian is conserved, we consider the 1-excitation subspace only. We consider the basis to be $|01111\rangle$, $|10111\rangle$, $|11011\rangle$, $|11101\rangle$ and $|11110\rangle$. This is unlike the spin chain case where we considered the basis to be of the form $|100\rangle$. The dynamics however are identical in both the cases. The Hamiltonian is 
    \begin{flalign}
    \mathcal{H}
    =\frac{1}{2}
    \begin{pmatrix}
    \lambda_1 & -J_{12} & 0 & 0 & 0 \\
    -J_{12} & \lambda_2 & -J_{23} & 0 & 0 \\
    0 & -J_{23} & \lambda_3 & -J_{34} & -J_{35} \\
    0 & 0 & -J_{34} & \lambda_4 & 0 \\
    0 & 0 & -J_{35} & 0 & \lambda_5
    \end{pmatrix} &&
    \end{flalign}
    where
    \begin{flalign}
    \begin{split}
    \lambda_1 =& \; h_1-h_2-h_3-h_4-h_5 \\
    \lambda_2 =& -h_1+h_2-h_3-h_4-h_5 \\
    \lambda_3 =& -h_1-h_2+h_3-h_4-h_5 \\
    \lambda_4 =& -h_1-h_2-h_3+h_4-h_5 \\
    \lambda_5 =& -h_1-h_2-h_3-h_4+h_5
    \end{split}    
     &&
    \end{flalign}
    \subsection{Routing Conditions}
    We perform a change of basis given by
    \begin{flalign}
    |10111\rangle &\rightarrow |+\rangle = |1\rangle\left(\frac{|10\rangle + |01\rangle}{\sqrt{2}}\right)|11\rangle \\
    |11011\rangle &\rightarrow |-\rangle = |1\rangle\left(\frac{|10\rangle - |01\rangle}{\sqrt{2}}\right)|11\rangle &&
    \end{flalign}
    The idea is to use two intermediate states that are resonant with the two output ports. This creates two routes for the input state to go through. Selecting one route is equivalent to making the input port resonant with that route. In this case, the two states $|+\rangle$ and $|-\rangle$ serve as the intermediate states through which state transport occurs.    
    Under the transformation, the Hamiltonian is given by
    \begin{flalign}
    \mathcal{H}
    =\frac{1}{2}
    \begin{pmatrix}
    \lambda_1 & \frac{-J_{12}}{\sqrt{2}} & \frac{-J_{12}}{\sqrt{2}} & 0 & 0 \\
    \frac{-J_{12}}{\sqrt{2}} & -h_1-h_4-h_5-J_{23} & h_2-h_3 & \frac{-J_{34}}{\sqrt{2}} & \frac{-J_{35}}{\sqrt{2}} \\
    \frac{-J_{12}}{\sqrt{2}} & h_2-h_3 & -h_1-h_4-h_5+J_{23} & \frac{J_{34}}{\sqrt{2}} & \frac{J_{35}}{\sqrt{2}} \\
    0 & \frac{-J_{34}}{\sqrt{2}} & \frac{J_{34}}{\sqrt{2}} & \lambda_4 & 0 \\
    0 & \frac{-J_{35}}{\sqrt{2}} & \frac{J_{35}}{\sqrt{2}} & 0 & \lambda_5
    \end{pmatrix} &&
    \end{flalign}
    Putting a condition that $h_2 = h_3$, all off-diagonal entries in the Hamiltonian are dependent on $J_{12}, J_{34}, J_{35}$ which are the non-gate couplings. If we assume that these couplings are much smaller than the diagonal entries, then the Hamiltonian is effectively diagonal. The (effective) eigenvalues are given by
    \begin{flalign}
    \begin{split}
&E_I = \frac{1}{2}\left(h_1-h_2-h_3-h_4-h_5\right) \\
&E_+ = \frac{1}{2}\left(-h_1-h_4-h_5-J_{23}\right) \\
&E_- = \frac{1}{2}\left(-h_1-h_4-h_5+J_{23}\right) \\
&E_{O_1} = \frac{1}{2}\left(-h_1-h_2-h_3+h_4-h_5\right) \\
&E_{O_2} = \frac{1}{2}\left(-h_1-h_2-h_3-h_4+h_5\right)
    \end{split} &&
    \end{flalign}
    To get state transfer, we need resonance between input port $(I)$ and output port $(O_1 \text{ or } O_2)$ through the intermediate states $\left |+\rangle \text{ or } |-\rangle\right)$. We consider the following scenario
    \begin{flalign}
    \notag
    I &\rightarrow |+\rangle \rightarrow O_1 \\
    \notag
    I &\rightarrow |-\rangle \rightarrow O_2 &&
    \end{flalign}
    Consider the first case. The conditions obtained by equating their energies are
    \begin{flalign}
    \begin{split}
    \label{cd1}
&h_1 = h_4 \\
&h_2-h_1 = \frac{J_{23}}{2}
    \end{split} &&
    \end{flalign}
    and similarly for the second case, the conditions obtained are
        \begin{flalign}
    \begin{split}
    \label{cd2}
&h_1 = h_5 \\
&h_1-h_2 = \frac{J_{23}}{2}
    \end{split} &&
    \end{flalign}
    These conditions lead to state transfer between the input and output ports for some time $\tau$. We determine $\tau$ by acting the propagator on the input state and equating it to the output state. We will consider a simple case by reducing the independent parameters. Consider state transfer from $I$ to $O_1$. We impose the following conditions on the parameters which satisfy the conditions obtained in equation \ref{cd1}
    \begin{flalign}
    \begin{split}
&J_{12} = J_{34} = J_{35} = J \\
&J_{23} = G \\
&h_2 = h_3 = 0 \\
&h_1 = h_4 = h_2 - \frac{J_{23}}{2} = -\frac{G}{2} \\
&h_5 = -h_1 = \frac{G}{2}
    \end{split} &&
    \end{flalign}
    Similar conditions are obtained for the second case. The conditions are summarized in the table given below
\begin{table}[H]
\centering
\caption{Transport Parameters for 5-Spin Router}
\label{params}
\begin{tabular}{|l|l|l|l|l|l|l|l|l|l|}
\hline
               & \textbf{$h_1$} & \textbf{$h_2$} & \textbf{$h_3$} & \textbf{$h_4$} & \textbf{$h_5$} & \textbf{$J_{12}$} & \textbf{$J_{23}$} & \textbf{$J_{34}$} & \textbf{$J_{35}$} \\ \hline
\textbf{$O_1$} & $\frac{-G}{2}$ & $0$            & $0$            & $\frac{-G}{2}$ & $\frac{G}{2}$  & $J$               & $G$               & $J$               & $J$               \\ \hline
\textbf{$O_2$} & $\frac{G}{2}$  & $0$            & $0$            & $\frac{-G}{2}$ & $\frac{G}{2}$  & $J$               & $G$               & $J$               & $J$               \\ \hline
\end{tabular}
\end{table}
\noindent
All the parameters except $h_1$ are constant for the two cases. To achieve routing of information, we thus require control over only the input port of the chain.
    To calculate $\tau$, consider the Hamiltonian for the first case 
    \begin{flalign}
    \mathcal{H}
    =
    \begin{pmatrix}
    -x & -y & -y & 0 & 0 \\
    -y & -x & 0 & -y & -y \\
    -y & 0 & 3x & y & y \\
    0 & -y & y & -x & 0 \\
    0 & -y & y & 0 & 3x
    \end{pmatrix} &&
    \end{flalign}
    where $x = \frac{G}{4}$ and $y = \frac{J}{2\sqrt{2}}$
    
    The states $|-\rangle$ and $|O_2\rangle$ are completely off-resonance with difference in energies being $|4x| = |G|$.
    We will only consider the reduced Hamiltonian comprising of the input state $|I\rangle$, the intermediate state $|+\rangle$ and the output state $|O_1\rangle$ to get an estimate of $\tau$. 
    \begin{flalign}
    \mathcal{H}
    =
    \begin{pmatrix}
    -x & -y & 0 \\
    -y & -x &-y \\
    0 & -y & -x
    \end{pmatrix}
    =
    \begin{pmatrix}
    -\frac{G}{4} & -\frac{J}{2\sqrt{2}} & 0 \\
    -\frac{J}{2\sqrt{2}} & -\frac{G}{4} & -\frac{J}{2\sqrt{2}} \\
    0 & -\frac{J}{2\sqrt{2}} & -\frac{G}{4}
    \end{pmatrix}    
     &&
    \end{flalign}
    The eigenvalues of this Hamiltonian are 
    \begin{flalign}
    \begin{split}
    &\lambda_1 = -\frac{G}{4} \\
    &\lambda_2 = -\frac{G}{4} - \frac{J}{2} \\
    &\lambda_3 = -\frac{G}{4} + \frac{J}{2}
    \end{split} &&
    \end{flalign}
    and the corresponding (non-normalized) eigenvectors are
    \begin{flalign}
    v_1
    =
    \begin{pmatrix}
    -1 \\
    0 \\
    1
    \end{pmatrix}
    \text{,  }
    v_2
    =
    \begin{pmatrix}
    1 \\
    -\sqrt{2} \\
    1
    \end{pmatrix}
    \text{,  }
    v_3
    =
    \begin{pmatrix}
    1 \\
    \sqrt{2} \\
    1
    \end{pmatrix}
    &&
    \end{flalign}
    Since we are considering state transfer from $I$ to $O_1$, the initial and final states are 
    \begin{flalign}
    \begin{split}
    &\text{Initial State : } \begin{pmatrix}
    1 \\
    0 \\
    0
    \end{pmatrix}     \\
    &\text{After $t = \tau$, Final State : }\begin{pmatrix}
    0 \\
    0 \\
    1
    \end{pmatrix}
    \end{split} &&
    \end{flalign}
    We write the initial state in terms of the eigenvectors of the Hamiltonian to calculate $\tau$.
    \begin{flalign}
    \begin{pmatrix}
    1 \\
    0 \\
    0
    \end{pmatrix}
    =
    \left(\frac{-1}{2}\right)v_1 + \left(\frac{1}{4}\right)v_2 + \left(\frac{1}{4}\right)v_3
    &&
    \end{flalign}
    To solve for $\tau$, we have
    \begin{flalign}
    e^{-i\mathcal{H}\tau} \begin{pmatrix}
    1 \\
    0 \\
    0
    \end{pmatrix}
    =
    \begin{pmatrix}
    0 \\
    0 \\
    1
    \end{pmatrix}
    &&
    \end{flalign}
    In terms of eigenvalues and eigenvectors of $\mathcal{H}$,
    \begin{flalign}
    \left(\frac{-1}{2}\right)e^{-i\lambda_1\tau}v_1 +
    \left(\frac{1}{4}\right)e^{-i\lambda_2\tau}v_2 + 
    \left(\frac{1}{4}\right)e^{-i\lambda_3\tau}v_3
    = \begin{pmatrix}
    0 \\
    0 \\
    1
    \end{pmatrix}
    &&
    \end{flalign}
    Solving for $\tau$, we get the following conditions
    \begin{flalign}
    \begin{split}
&e^{-i\lambda_1\tau} = -1 \\
&e^{-i\lambda_2\tau} = 1 \\
&e^{-i\lambda_3\tau} = 1
    \end{split} &&
    \end{flalign}
    which can be written as
    \begin{flalign}
    \begin{split}
&\left(\frac{G}{4}\right)\tau = (2n_1+1)\pi \\
&\left(\frac{G}{4}+\frac{J}{2}\right)\tau = 2n_2\pi \\
&\left(\frac{G}{4}-\frac{J}{2}\right)\tau = 2n_3\pi
    \end{split} &&
    \end{flalign}
    These conditions can be further reduced to
    \begin{flalign}
    \begin{split}
&G\tau = 4(m_1+m_2)\pi \\
&J\tau = 2(m_1-m_2)\pi
    \end{split} &&
    \end{flalign}
    This puts constraints on the values taken by integers $m_1$ and $m_2$. In terms of $G$ and $J$, $m_2$ is given by
    \begin{flalign}
    m_2 = \left(\frac{G-2J}{G+2J}\right)m_1 &&
    \end{flalign}
    and the time is given by
    \begin{flalign}
    \label{RoutingTime}
    \tau = \left(\frac{8m_1}{G+2J}\right)\pi &&
    \end{flalign}
    For example, if $G=2\pi*100$ and $J=2\pi*10$ then
    \begin{flalign}
    m_2 = \left(\frac{100-20}{100+20}\right)m_1=\left(\frac{80}{120}\right)m_1=\left(\frac{2}{3}\right)m_1 &&
    \end{flalign}
    Thus $m_1$ can only take values 3,6,9,... The smallest possible time is then for $m_1=3$.
      \begin{flalign}
    \tau_{min} = 0.1 &&
    \end{flalign}
    \subsection{Simulations}
    We simulate the system with the parameters used above. As before, we consider two input states : $|1\rangle$ and $|+\rangle$.
    \begin{figure}[H]
    \centering
            \def\svgwidth{0.8\linewidth}
		    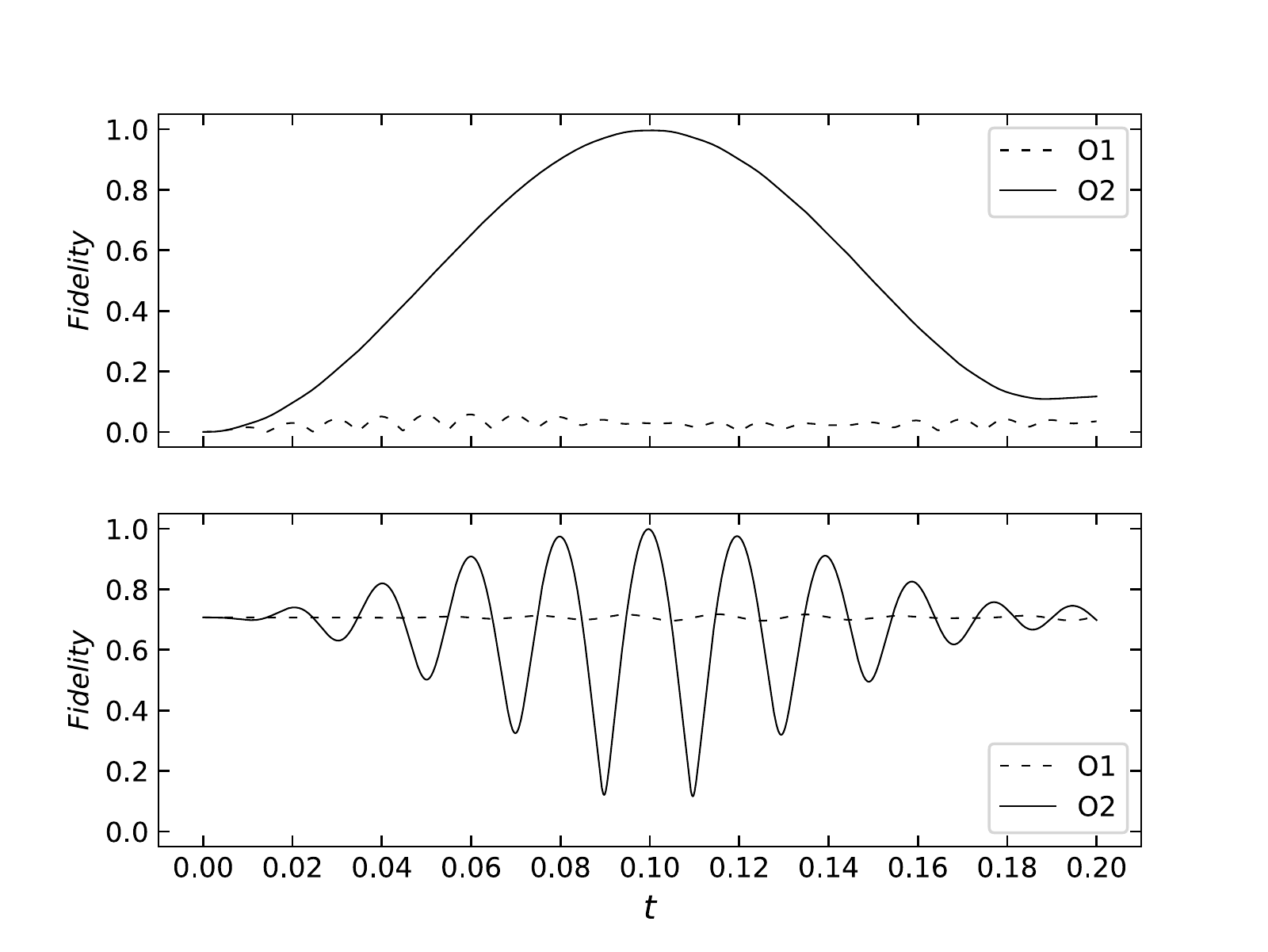
		    \caption{5-Spin Router - Fidelity vs $t$ for normal case}
		    \label{fig:router5}
\end{figure}
\noindent
We expect peak fidelity at $t=0.1$ while the superposition transport peak fidelity is slightly less than $1$. This is justified since the expected time is calculated approximately. Fidelity profile for superposition transport is much more robust in terms of optimal time unlike that of the 4-spin system where rapid oscillations in fidelity were observed.
\begin{figure}[H]
\centering
            \def\svgwidth{0.8\linewidth}
		    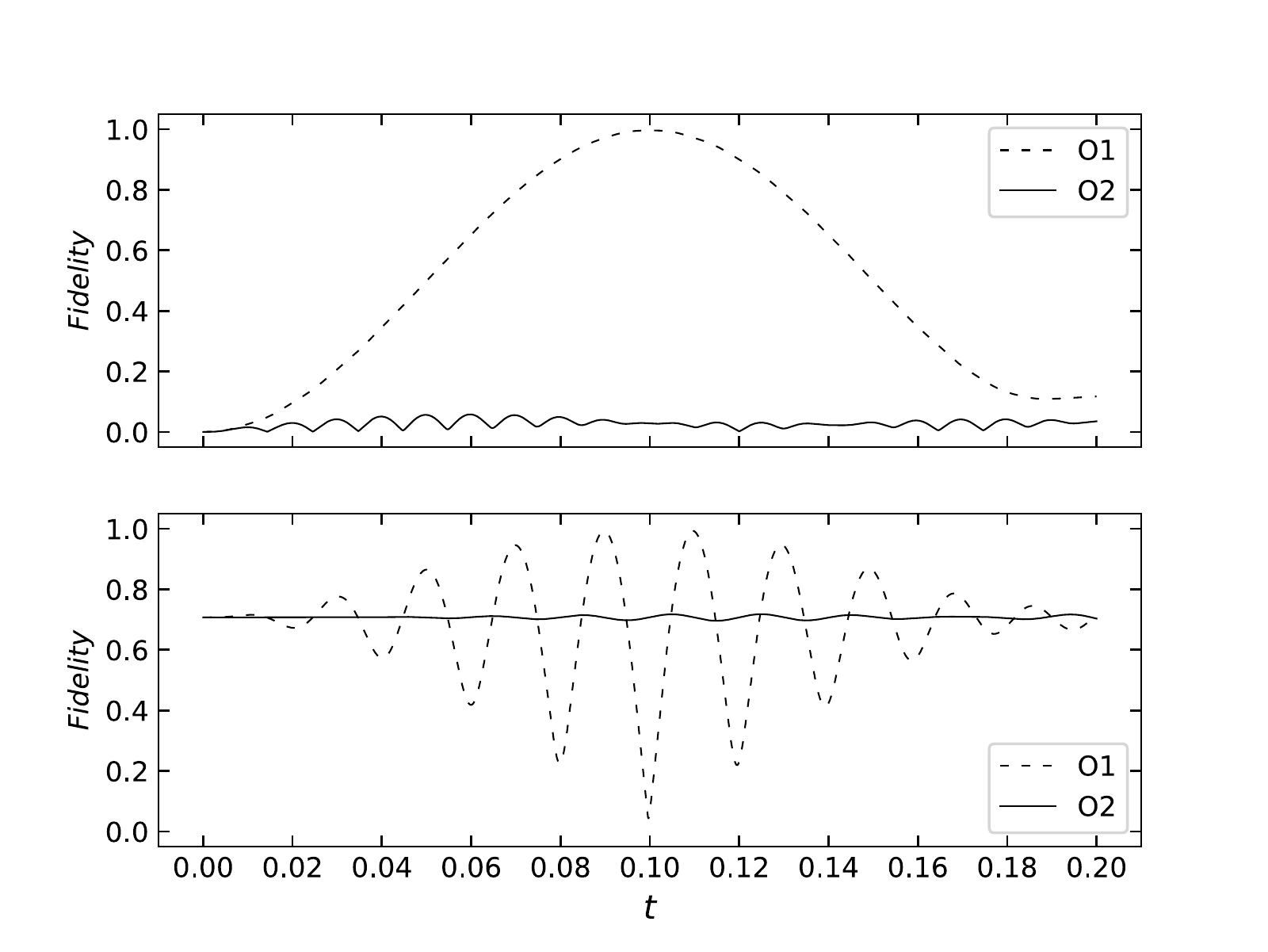
		    \caption{5-Spin Router - Fidelity vs $t$ for switched case}
		    \label{fig:router5_switched}
\end{figure}
\noindent
Unlike the first case, when we switch the magnetic field on the input spin, the fidelity for superposition transport at $t=0.1$ is almost $0$. The resulting state is the corresponding orthogonal state $\left(|0\rangle-|1\rangle\right)/2$. A phase factor of $\pi$ is picked up in this case which can be overcome by applying a Z-Gate to the final state. Alternatively, we can move the control aspect from the input node to the output nodes. If we want to switch the route, we switch the magnetic fields of the output nodes. Effectively by moving the control to the output nodes, both the routing scenarios obey the first case observed in Figure \ref{fig:router5}.

    \section{Modularity}
    \begin{figure}[H]
\centering
\begin{tikzpicture}
\begin{scope}[every node/.style={circle,thick,draw}]
    \node (1) at (0,0) {1};
    \node (2) at (2,0) {2};
    \node (3) at (4,0) {3};
    \node (4) at (6,0) {4};
    \node (5) at (7.5,1.5) {5};
    \node (6) at (7.5,-1.5) {6};
\end{scope}

\begin{scope}[>={Stealth[black]},
              every node/.style={fill=white,circle},
              every edge/.style={draw=black,thick}]
    \path [-] (1) edge (2);
    \path [-] (2) edge (3);
    \path [-] (3) edge (4);
    \path [-] (4) edge (5);
    \path [-] (4) edge (6);
\end{scope}
\end{tikzpicture}
\label{fig:img_modularity}
\caption{Modular combination of 3-spin chain and 4-spin router}
\end{figure}
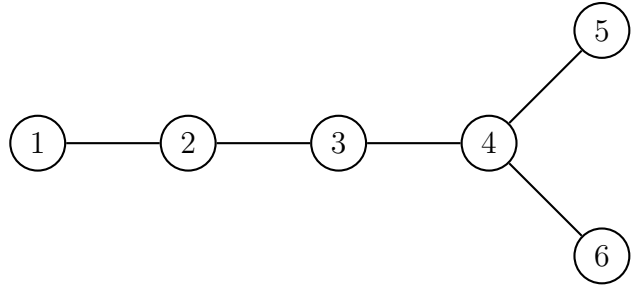
\begin{figure}[H]
\centering
            \def\svgwidth{0.8\linewidth}
		    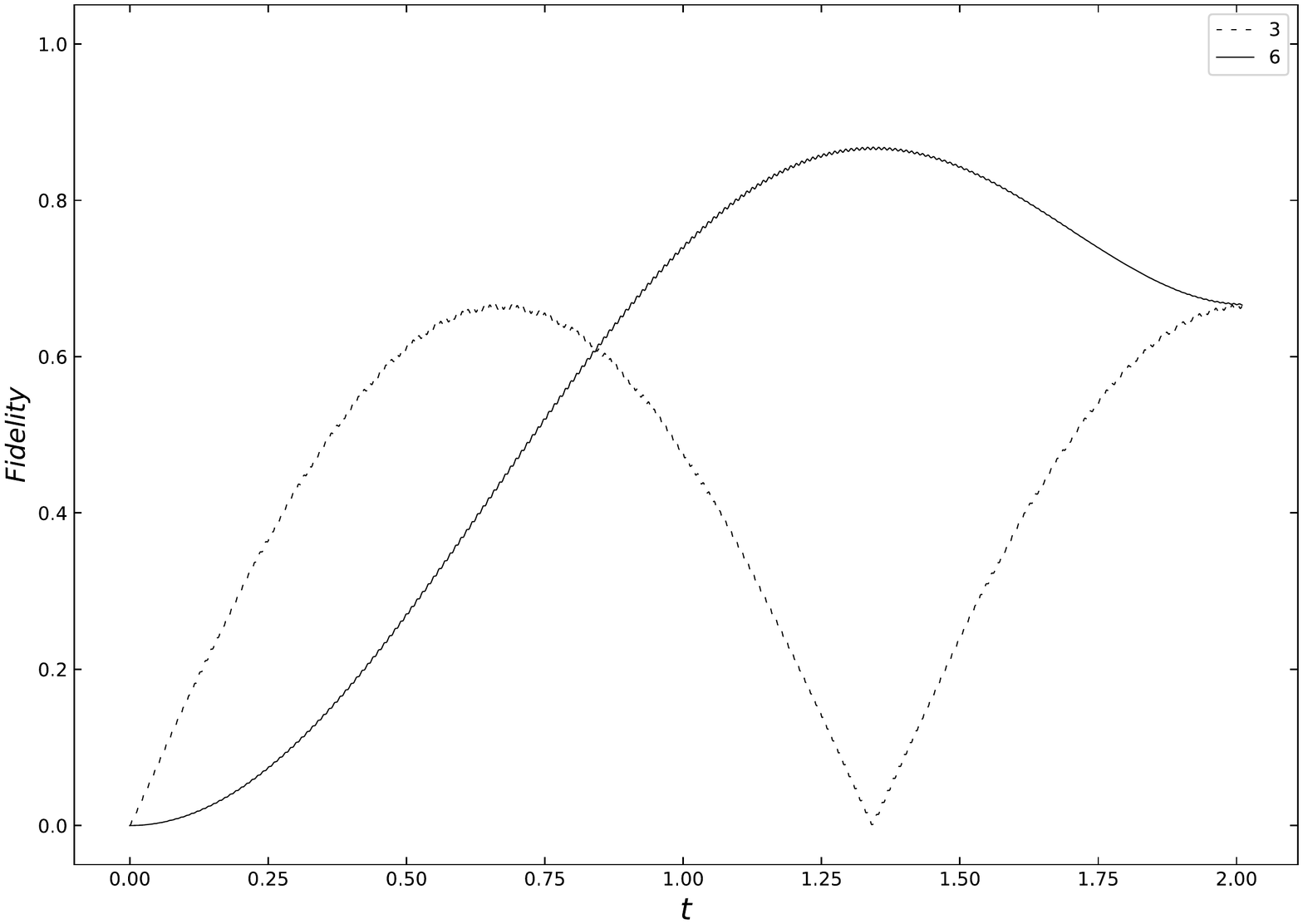
		    \caption{Fidelity vs $t$ for naive modular combination}
		    \label{fig:modular_naive}
\end{figure}

We have shown two simple blocks of networks that achieve some transport purpose. Similar blocks can thus be constructed. These blocks of networks can provide different applications in a computer. If one needs to combine these, then a straightforward combination of the blocks does not achieve the purpose. The second block interferes with working of the first block and vice-versa. Consider a naive combination of a 3-spin chain and a 4-spin router. Figure \ref{fig:modular_naive} shows the dynamics of such a combination.
\noindent 
While the transport from one end of the chain to other is happening, there is leakage to the router which can be seen in the fidelity of the final ($6^{th}$) spin the system. This can be overcome by switching off the blocks not in use in some way. A strong magnetic field on the spin adjacent to the block in use achieves this. It acts as a barrier that prevents the first block from communicating with the second one. We apply strong magnetic fields on the spin adjacent to the chain. Once the dynamics of the chain is complete, we remove this field and apply it on the spin adjacent to the router. Such a mechanism creates a barrier and allows transport from the chain to the router without leakage. Consider a time dependent magnetic field on spins 2 and 4. This results in the following dynamics
\begin{figure}[H]
\centering
            \def\svgwidth{0.8\linewidth}
		    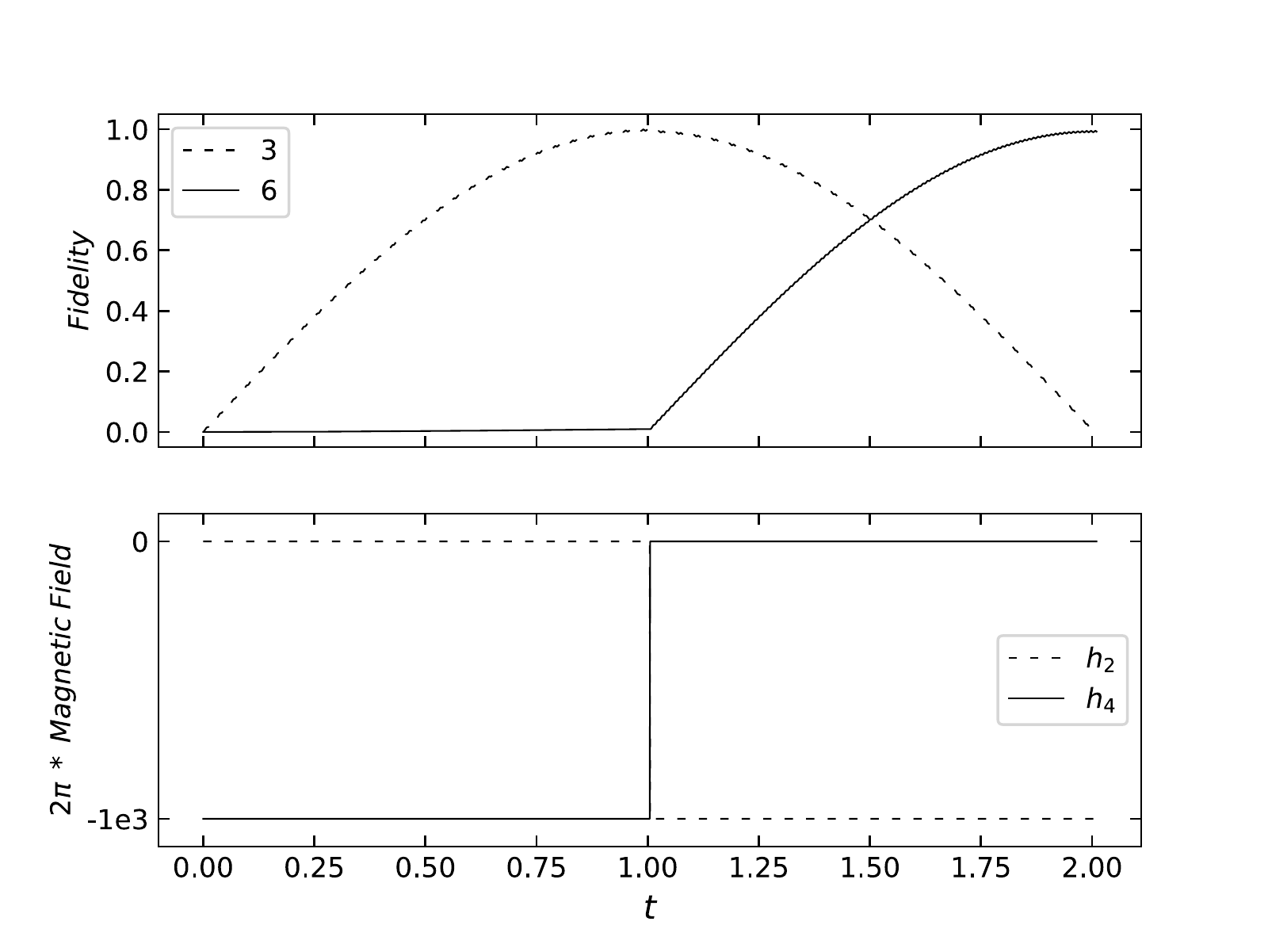
		    \caption{Fidelity vs $t$ for modular combination with time-dependent magnetic fields}
		    \label{fig:modular}
\end{figure}
\newpage
    \section{Network}    
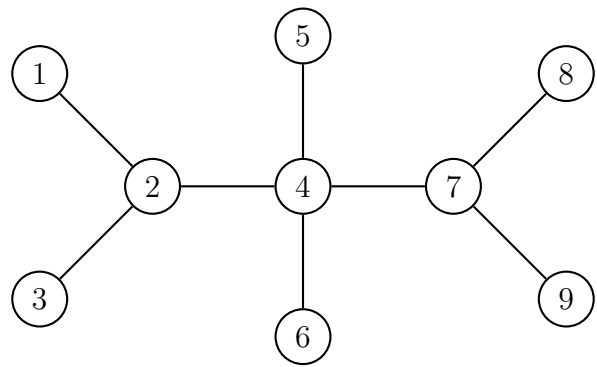
\begin{figure}[H]
\centering
\begin{tikzpicture}
\begin{scope}[every node/.style={circle,thick,draw}]
    \node (1) at (-1.5,1.5) {1};
    \node (2) at (0,0) {2};
    \node (3) at (-1.5,-1.5) {3};
    \node (4) at (2,0) {4};
    \node (5) at (2,2) {5};
    \node (6) at (2,-2) {6} ;
    \node (7) at (4,0) {7} ;
    \node (8) at (5.5,1.5) {8} ;
    \node (9) at (5.5,-1.5) {9} ;
\end{scope}

\begin{scope}[>={Stealth[black]},
              every node/.style={fill=white,circle},
              every edge/.style={draw=black,thick}]
    \path [-] (1) edge (2);
    \path [-] (2) edge (3);
    \path [-] (2) edge (4);
    \path [-] (4) edge (5);
    \path [-] (4) edge (6);
    \path [-] (4) edge (7);
    \path [-] (7) edge (8);
    \path [-] (7) edge (9); 
\end{scope}
\end{tikzpicture}
\caption{Spin network in arbitrary topology}
\label{fig:network_other}
\end{figure}

\begin{figure}[H]
\centering
            \def\svgwidth{0.8\linewidth}
		    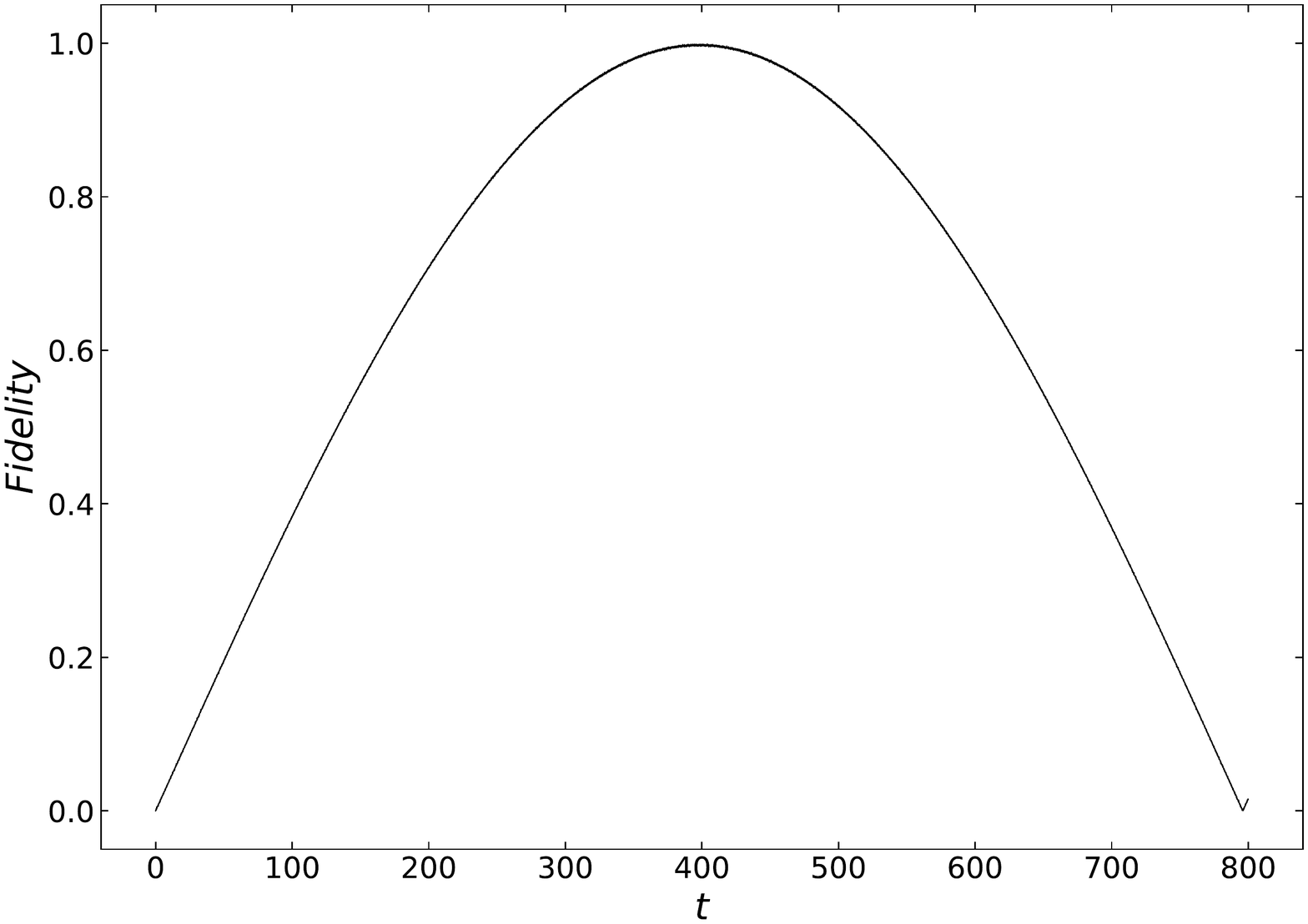
		    \caption{Fidelity vs $t$ in arbitrary topology network}
		    \label{fig:network_new}
\end{figure}

    In previous work, we discussed information transport in a structured spin system. Here we consider a general network and comment on the possibility of transport from one node to another. We consider small systems for ease of simulation. Similar extensions can be made to larger or more complex topologies. 
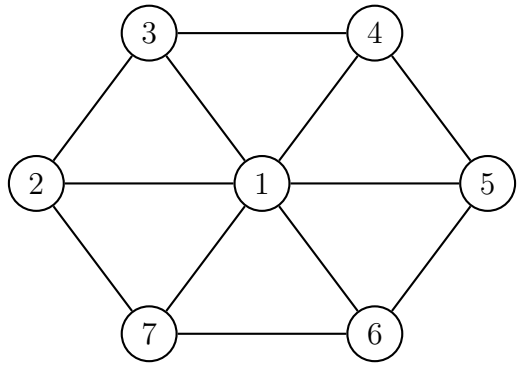
\begin{figure}[H]
\centering
\begin{tikzpicture}
\begin{scope}[every node/.style={circle,thick,draw}]
    \node (1) at (0,0) {1};
    \node (2) at (-3,0) {2};
    \node (3) at (-1.5,2) {3};
    \node (4) at (1.5,2) {4};
    \node (5) at (3,0) {5};
    \node (6) at (1.5,-2) {6} ;
    \node (7) at (-1.5,-2) {7} ;
\end{scope}

\begin{scope}[>={Stealth[black]},
              every node/.style={fill=white,circle},
              every edge/.style={draw=black,thick}]
    \path [-] (1) edge (2);
    \path [-] (1) edge (3);
    \path [-] (1) edge (4);
    \path [-] (1) edge (5);
    \path [-] (1) edge (6);
    \path [-] (1) edge (7);
    \path [-] (2) edge (3);
    \path [-] (3) edge (4);
    \path [-] (4) edge (5); 
    \path [-] (5) edge (6); 
    \path [-] (6) edge (7); 
    \path [-] (7) edge (2); 
\end{scope}
\end{tikzpicture}
\caption{Spin network in wheel topology}
\label{fig:network_wheel}
\end{figure}
\noindent
Consider a wheel topology with one central spin and six peripheral spins as shown in Figure \ref{fig:network_wheel}. We attempt to transport information from one node on the ring to the opposite node on the ring based on the resonance phenomenon (for example from node 2 to node 5). The couplings $(2\pi*10)$ are assumed to be constant as before while there is magnetic field $(2\pi*100)$ only on the input and output nodes. Since this a much more complex topology than a chain, there is a possibility of dispersion of information through the network. However, we still obtain peak fidelity on the target spin at some time.
\begin{figure}[H]
\centering
            \def\svgwidth{0.8\linewidth}
		    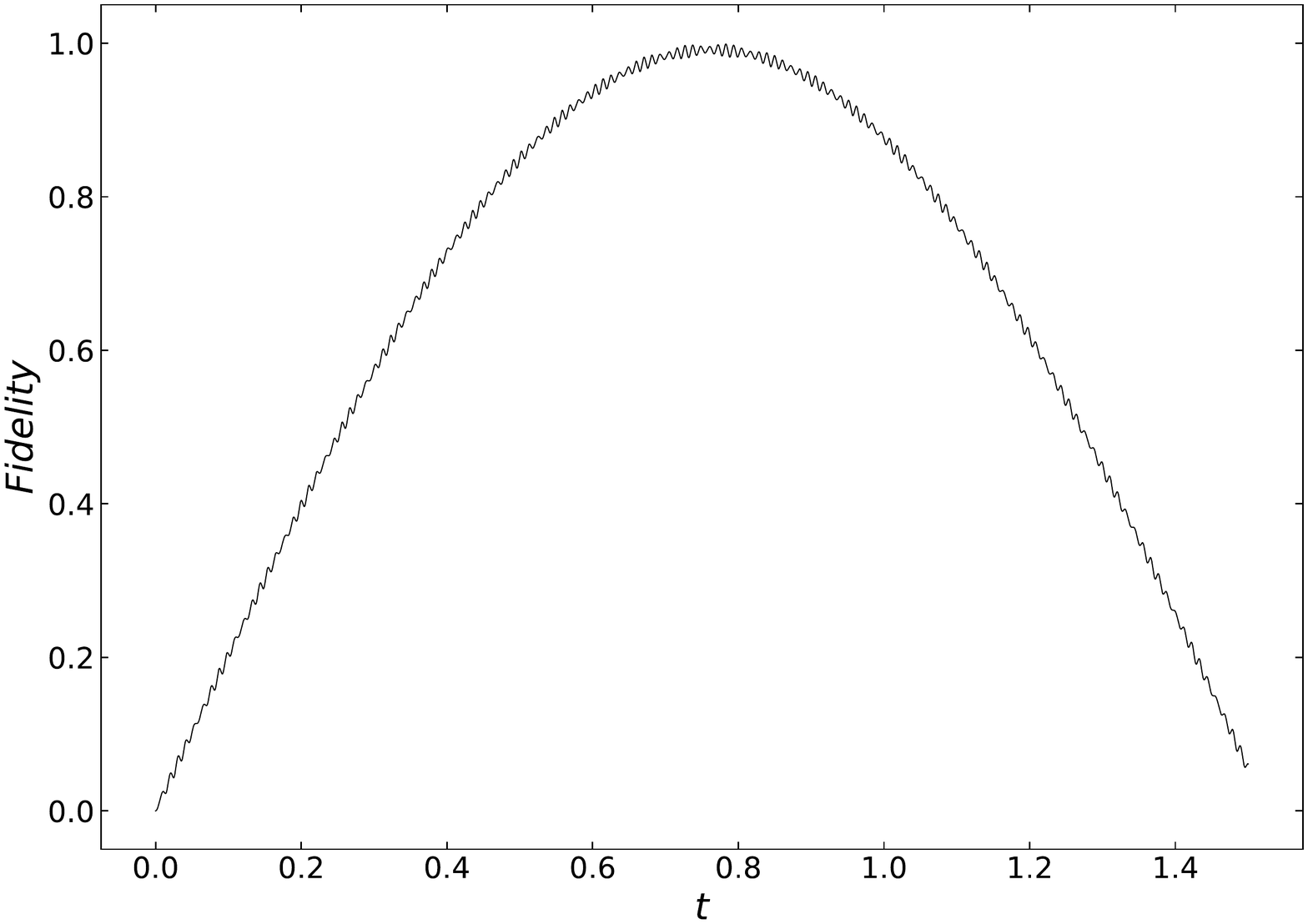
		    \caption{Fidelity vs $t$ in wheel topology network}
		    \label{fig:network}
\end{figure}
\noindent
We consider another topology given in Figure \ref{fig:network_other}. We impose the same parameters on the system. We attempt to transport information from node 1 to node 9 of the network. The fidelity profile for the target spin is given in Figure \ref{fig:network_new}. 
\noindent
While even this network permits information transport, the time taken for this is quite large. However, this can be controlled by changing the coupling and magnetic field parameters. This opens up much more avenues of transport in any arbitrary topology of networks.

\section{Conclusion}
We discussed quantum information transport in various types of spin network structures. To minimize the the introduction of noise in the protocol, we kept the parameters of the model as non-invasive as possible. In most of the models discussed, we had uniform couplings with magnetic fields only on the input and output nodes. We also discussed analytical solution to the routing problem based on mechanism of a controlling gate comprising of the two central spins.

All the protocols discussed are efficient in the transport of information. We studied the robustness of the parameters of the chain model as a benchmark for other models. The modularity scheme discussed provides a valuable solution to combining protocols but introduces additional noise by requiring control on specific spins. Finally, we discussed arbitrary network transport based on the resonance phenomenon. 

\chapter{Discussion and Conclusion}
We discussed two methodologies based on the framework of spin networks - 
\begin{itemize}
\item \textbf{Star topology engineering} \\
Using the filtered Hamiltonian engineering approach, in a star topology with arbitrary connections, we decoupled the peripheral-peripheral interactions and retained the radial interactions to create a perfect star with only radial interactions. We showed analytically and verified by simulation that such a scheme would work for every fourth cycle using a grating or filter function. The filter function is the key in this technique since it appears in every interaction of the average Hamiltonian and thus we can set its argument to a value that would decouple or retain that interaction. 
\item \textbf{Information transport} \\
Information transport in spin networks is a vast field where lot of models have been proposed previously. In this work, we proposed a linear chain and a router for transport. We showed that such a scheme can be implemented with simple parameters using the phenomenon of resonance. The technique is robust in terms of the parameters but has a short window of peak fidelity. This is generally not a problem since the measurement devices typically are capable of operating in much smaller time scales. We also discussed a modularity scheme where two protocols can be combined with the help of time-dependent magnetic fields. This achieves its purpose at the cost of extra invasive control on specific spins. Since these protocols were based on specific structures of the network, we considered a general network to demonstrate the resonance phenomenon. Working out the details of any general network is a very hard problem. Instead we demonstrated transport in two different topologies of network.
\end{itemize}
	
\chapter{Outlook and Future Plans}
The work done in the thesis was based on the architecture of spin networks. Spin networks are a valuable test-bed to study interesting theoretical problems such as decision tree problem \cite{FarhiDecTree}, quantum random walks \cite{QuantumRandomWalks}. Forays can be made if spin networks can be used for analog quantum simulation by mapping it to a different model. Christandl et. al. showed in \cite{QTSchemes1} that in the XY model, spin $\frac{1}{2}$ particles can be mapped to spinless fermions. Such techniques can be used for quantum simulation.

The discussion of Hamiltonian engineering pertained to the creation of star topology only. The technique of filtered Hamiltonian engineering can be extended to create different topologies as well. It does not provide an improvement in terms of individual control on indistinguishable spins but it does give an added control on selectivity as discussed in Section \ref{sec:star_concl}. In a broader sense, this technique has applications in quantum simulation as well, as shown in \cite{AjoyHamEng}. Assuming individual accessibility of spins, this technique can very useful for generation of different types of propagators.

The information transport protocols discussed in this work provide alternative solutions to the problem of transport using natural evolution of the system. We discussed the presence of a resonance phenomenon which is not established rigorously in this work. Finding the cause of this would provide a base to construct different transport models. In the models discussed, a degree of control on individual spins is still required despite keeping minimal requirements on the parameters of the model. Working out a model that permits transport for arbitrary couplings would be a marked improvement on the existing models.

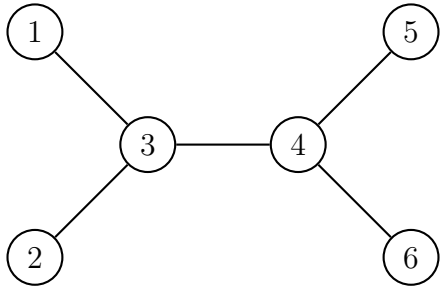
\begin{figure}[H]
\centering
\begin{tikzpicture}
\begin{scope}[every node/.style={circle,thick,draw}]
    \node (1) at (0,1.5) {1};
    \node (2) at (0,-1.5) {2};
    \node (3) at (1.5,0) {3};
    \node (4) at (3.5,0) {4};
    \node (5) at (5,1.5) {5};
    \node (6) at (5,-1.5) {6};
\end{scope}

\begin{scope}[>={Stealth[black]},
              every node/.style={fill=white,circle},
              every edge/.style={draw=black,thick}]
    \path [-] (1) edge (3);
    \path [-] (2) edge (3);
    \path [-] (3) edge (4);
    \path [-] (4) edge (5);
    \path [-] (4) edge (6);
\end{scope}
\end{tikzpicture}
\caption{Spin network architecture for CNOT gate}
\label{fig:img_cnot}
\end{figure}

Information transport protocols in spin networks can also be extended as a mechanism to obtain logic gates. We mentioned this possibility in the thesis summary \ref{summary} but have not commented on it yet. A simple example of this is the SWAP gate which swaps two input qubits based on the control qubit. This gate can be implemented in an information transport framework by having two input spins $(I_1,I_2)$ and two output spins $(O_1,O_2)$. The control spin will determine if information flow will happen in straightforward manner $I_1\rightarrow O_1$ and $I_2\rightarrow O_2$ or the opposite. These two cases constitute the SWAP gate. Such construction of gates is however limited by the fact that the XY Hamiltonian is spin number preserving. If we want to construct a CNOT gate using information transport, this is not possible since CNOT gate changes the total spin number. An alternative to this is using logical qubits so that the total spin is conserved in each case of CNOT truth table. The truth tables for CNOT gate with standard qubits is given below. The two tables correspond to the initial and final state of the network.
\begin{table}[H]
\centering
\caption{CNOT Truth Table}
\label{cnot_tt}
\begin{tabular}{|l|l|l|l|l|l|l|}
\cline{1-3} \cline{5-7}
\textbf{Control} & \textbf{Input} & \textbf{Output} & \textbf{}   & \textbf{Control} & \textbf{Input} & \textbf{Output} \\ \cline{1-3} \cline{5-7} 
0          & 0          & 0          &             & 0          & 0          & 0          \\ \cline{1-3} \cline{5-7} 
0          & 1          & 0          & $\rightarrow$ & 0          & 1          & 1          \\ \cline{1-3} \cline{5-7} 
1          & 0          & 0          &             & 1          & 0          & 1          \\ \cline{1-3} \cline{5-7} 
1          & 1          & 0          &             & 1          & 1          & 0          \\ \cline{1-3} \cline{5-7} 
\end{tabular}
\end{table}
\noindent
The quantum number is not conserved in case of standard qubits. However, if we consider logical qubits with the mapping
$|0\rangle\rightarrow|01\rangle$ and $|1\rangle\rightarrow|10\rangle$, then the quantum number is conserved and such evolution is allowed in the transport framework.
\begin{table}[H]
\centering
\caption{CNOT Truth Table (Logical Qubits)}
\label{cnot_tt_logical}
\begin{tabular}{|l|l|l|l|l|l|l|}
\cline{1-3} \cline{5-7}
\textbf{Control} & \textbf{Input} & \textbf{Output} & \textbf{}   & \textbf{Control} & \textbf{Input} & \textbf{Output} \\ \cline{1-3} \cline{5-7} 
01          & 01          & 01          &             & 01          & 01          & 01          \\ \cline{1-3} \cline{5-7} 
01          & 10          & 01          & $\rightarrow$ & 01          & 10          & 10          \\ \cline{1-3} \cline{5-7} 
10          & 01          & 01          &             & 10          & 01          & 10          \\ \cline{1-3} \cline{5-7} 
10          & 10         & 01          &             & 10          & 10          & 01          \\ \cline{1-3} \cline{5-7} 
\end{tabular}
\end{table}
\noindent
To confirm the feasibility of this scheme, we use the optimization routine of MATLAB's genetic algorithm to find a set of parameters (including time) that result in a CNOT gate. Concretely, we consider the four input states $|I_1\rangle = |010101\rangle$, $|I_2\rangle = |011001\rangle$, $|I_3\rangle = |100101\rangle$ and $|I_4\rangle = |101001\rangle$ and evolve them under the XY Hamiltonian with constant couplings and variable magnetic fields. We consider the architecture in Figure \ref{fig:img_cnot} where the spins 1 and 2 are the input, 3 and 4 are the control and 5 and 6 are the output.
\noindent
Considering constant coupling $J$, we optimize for the 8 parameters ($J$, 6 Magnetic fields, $t$) for the cost function $\frac{1}{4}\left|\sum_i 1 -\langle O_i|U|I_i\rangle\right|$ where the output states are $|O_1\rangle = |010101\rangle$, $|O_2\rangle = |011010\rangle$, $|O_3\rangle = |100110\rangle$ and $|O_4\rangle = |101001\rangle$. We found the following set of parameters that result in the cost function value of $0.0111$.
\begin{flalign}
\begin{split}
&J = -78.2278 \\
&h = [304.2089, 58.5906, -749.6377, 196.3780, 64.4191, 61.9356] \\
&t = 30.9105
\end{split}
&&
\end{flalign}
To verify the scheme, we consider an equal superposition of all the input states. At time $t=30.9105$, the overlap of the output state with the target state obtained was $0.9868$. This confirms that the construction of logic gates in the information transport framework is quite feasible. The set of parameters could be further optimized or we could allow variable couplings for more degrees of freedom. Since analytical solutions exist in the case of a chain \cite{CappellaroChain}, logic gate solutions could be obtained by considering a linear architecture. An extension to this is to attach wires that transport information. Since CNOT can be used to entangle qubits, using this architecture one can obtain spatially separated entangled qubits. Although we are using logical qubits, it should be possible to go back to standard qubits using the techniques described by Kay and Ericsson \cite{KayGeometric}. Another interesting approach to construction of logic gates is to use the technique of natural evolution of the system to achieve a target unitary. Such a mechanism was demonstrated by DiVincenzo \cite{ExchangeComp} using exchange interactions. This area of creating logic gates using networks is very promising given that such gates can be combined in modular fashion to construct circuits.

\printbibliography		
\appendix
\chapter{Toggling Frame Hamiltonian}
\label{appendix:tfh}
We derive the toggling frame Hamiltonian defined in equation \ref{eqn:tfh} here.
\begin{flalign}
        U_Z\left(\tau\right)^\dagger U_{DQ}\left(t\right) U_Z\left(\tau\right) = \text{ exp}\left[{-i t \mathcal{H}_m\left(\tau\right)}\right] &&
        \end{flalign}        
We have
\begin{flalign}
\label{eqn:A1_diff0}
        \mathcal{H}_m = U_Z\left(\tau\right)^\dagger \mathcal{H}_{DQ}\left(t\right) U_Z\left(\tau\right) &&
        \end{flalign}
Taking derivative with respect to $\tau$,
\begin{flalign}
\label{eqn:A1_1}
        \frac{d\mathcal{H}_m}{d\tau} = -iU_Z\left(\tau\right)^{\dagger}\left[\mathcal{H}_{DQ},\mathcal{H}_Z\right]U_Z\left(\tau\right) &&
        \end{flalign}
The commutation relation $\left[\mathcal{H}_{DQ},\mathcal{H}_Z\right]$ is given by
\begin{flalign}
\label{eqn:A1_2}
\left[\mathcal{H}_{DQ},\mathcal{H}_Z\right]=\sum_{l,m}b_{lm}\left(S_l^xS_m^x-S_l^yS_m^y\right)\sum_n\omega_nS_n^z-\sum_n\omega_nS_n^z\sum_{l,m}b_{lm}\left(S_l^xS_m^x-S_l^yS_m^y\right)
&&
\end{flalign}
The spatial indices $l$ and $m$ are not to be confused with the subscript of the toggling frame Hamiltonian $\mathcal{H}_m$. Consider the first term in equation above,
\begin{flalign}
\label{eqn:A1_3}
\begin{split}
& \sum_{l,m,n}b_{lm}\omega_n\left(S_l^xS_m^xS_n^z-S_l^yS_m^yS_n^z\right) \\
&\text{This can be simplified using the spin commutation relations.} \\
=& \sum_{l,m,n}b_{lm}\omega_n\left(S_l^xS_n^zS_m^x+S_l^x\left[S_m^x,S_n^z\right]-S_l^yS_n^zS_m^y-S_l^y\left[S_m^y,S_n^z\right]\right) \\
=& \sum_{l,m,n}b_{lm}\omega_n\left(S_l^xS_n^zS_m^x-i\delta_{mn}S_l^xS_m^y-S_l^yS_n^zS_m^y-i\delta_{mn}S_l^yS_m^x\right) \\
=& \sum_{i,j,k}b_{ij}\omega_k\left(S_k^zS_i^xS_j^x+\left[S_l^x,S_n^z\right]S_m^x-S_n^zS_l^yS_m^y-\left[S_l^y,S_n^z\right]S_m^y-i\delta_{mn}S_l^xS_m^y-i\delta_{mn}S_l^yS_m^x\right)
\end{split}
&&
\end{flalign}
The first and third terms in the equation above cancel out with the second term of the commutation relation \ref{eqn:A1_2} and by working out the internal spin commutation relations, it simplifies to
\begin{flalign}
\left[\mathcal{H}_{DQ},\mathcal{H}_Z\right]=-i\sum_{l,m,n}b_{lm}\omega_n\left(\delta_{ln}S_l^yS_m^x+\delta_{ln}S_l^xS_m^y+\delta_{mn}S_l^xS_m^y+\delta_{mn}S_l^yS_m^x\right)
&&
\end{flalign}
We can lose one index by simplifying the Kronecker deltas. The commutation can be written as
\begin{flalign}
\left[\mathcal{H}_{DQ},\mathcal{H}_Z\right]=-i\sum_{l,m}b_{lm}(\omega_l+\omega_m)\left(S_l^xS_m^x+S_l^yS_m^y\right)
&&
\end{flalign}
By substituting the above commutation in equation \ref{eqn:A1_1}, we have
\begin{flalign}
\label{eqn:A1_diff1}
\frac{d\mathcal{H}_m}{d\tau}=-U_Z\left(\tau\right)\mathcal{H}_1U_Z\left(\tau\right)^{\dagger}
&&
\end{flalign}
where $\mathcal{H}_1 = \sum_{l,m}b_{lm}(\omega_l+\omega_m)\left(S_l^xS_m^x+S_l^yS_m^y\right)$

\noindent
Taking the second derivative of $\mathcal{H}_m$ with respect to $\tau$,
\begin{flalign}
\frac{d^2\mathcal{H}_m}{d\tau^2}=iU_Z\left(\tau\right)^{\dagger}\left[\mathcal{H}_1,\mathcal{H}_Z\right]U_Z\left(\tau\right)
&&
\end{flalign}
This is similar to equation \ref{eqn:A1_1}. We can work this out in a fashion similar to the methodology in equations \ref{eqn:A1_2} and \ref{eqn:A1_3}. We skip the details and give the final term of the second differential.
\begin{flalign}
\frac{d^2\mathcal{H}_m}{d\tau^2}=-U_Z\left(\tau\right)^{\dagger}\sum_{lm}b_{lm}(\omega_l+\omega_m)^2\left(S_l^xS_m^x-S_l^yS_m^y\right)
&&
\end{flalign}
Each individual term has the form of a standard Harmonic oscillator differential equation.
\begin{flalign}
\left(\frac{d^2\mathcal{H}_m}{d\tau^2}\right)_{lm}=-(\omega_l+\omega_m)^2\left(\mathcal{H}_m\right)_{lm}
&&
\end{flalign}
The solution to this is straightforward.
\begin{flalign}
\label{eqn:A1_4}
\left(\mathcal{H}_m(\tau)\right)_{lm} = Ae^{i(\omega_l+\omega_m)\tau}+Be^{-i(\omega_l+\omega_m)\tau}
&&
\end{flalign}
where $A$ and $B$ are terms to be determined by the conditions
\begin{flalign}
\begin{split}
&\left(\mathcal{H}_m(0)\right)_{lm} = \; A + B = \mathcal{H}_{DQ} \\
&\left(\frac{d\mathcal{H}_m}{d\tau}(0)\right)_{lm} = \; i(A-B)(\omega_l+\omega_m)=-\mathcal{H}_1
\end{split}
&&
\end{flalign}
The first equation is obtained by setting $\tau=0$ in equations \ref{eqn:A1_4} and \ref{eqn:A1_diff0} while the second one is obtained by setting $\tau=0$ in the derivative of equation \ref{eqn:A1_4} and equation \ref{eqn:A1_diff1}. Solving for $A$ and $B$ we get,
\begin{flalign}
\begin{split}
&A = \frac{b_{lm}}{2}S_l^-S_m^- \\
&B = \frac{b_{lm}}{2}S_l^+S_m^+
\end{split}
&&
\end{flalign}
where
\begin{flalign}
\begin{split}
&S_l^+= S_l^x + iS_l^y \\
&S_l^-=S_l^x-iS_l^y
\end{split}
&&
\end{flalign}
Substituting the values of $A$ and $B$ in equation \ref{eqn:A1_4}, the toggling frame Hamiltonian from is thus given by
\begin{flalign}
\mathcal{H}_m = \sum_{l,m}\frac{b_{lm}}{2}\left(S_l^-S_m^-e^{i(\omega_l+\omega_m)\tau} + S_l^+S_m^+e^{-i(\omega_l+\omega_m)\tau}\right)
&&
\end{flalign}
We further write $\omega_l+\omega_m$ as $\delta_{lm}$ in equation \ref{eqn:tfh}. This is not to be confused with the Kronecker delta.

\chapter{General Decoupling Conditions}
\label{appendix:gdc}
For a general sequence with $L$ stages, the total propagator is given by
\begin{flalign}
            \begin{split}
            U_N =& \left[U_Z^{(1_N 2_N 3_N \cdots {L-1}_N L_N)}\left(\tau\right)^\dagger U_{DQ}\left(\frac{t_L}{N}\right)  U_Z^{(1_N 2_N 3_N \cdots {L-1}_N L_N)}\left(\tau\right)\right] \cdot \\
            &\left[U_Z^{(1_N 2_N 3_N \cdots {L-1}_N L_{N-1})}\left(\tau\right)^\dagger U_{DQ}\left(\frac{t_{L-1}}{N}\right)  U_Z^{(1_N 2_N 3_N \cdots {L-1}_N L_{N-1})}\left(\tau\right)\right] \cdot \\
            &\left[U_Z^{(1_N 2_N 3_N \cdots {L-1}_{N-1} L_{N-1})}\left(\tau\right)^\dagger U_{DQ}\left(\frac{t_{L-2}}{N}\right)  U_Z^{(1_N 2_N 3_N \cdots {L-1}_{N-1} L_{N-1})}\left(\tau\right)\right] \cdot \\
            & \qquad \qquad \qquad \vdots \\
            &\left[U_Z^{(1_{N-1} 2_{N-1} 3_{N-1} \cdots {L-1}_{N-1} L_{N-1})}\left(\tau\right)^\dagger U_{DQ}\left(\frac{t_L}{N}\right)  U_Z^{(1_{N-1} 2_{N-1} 3_{N-1} \cdots {L-1}_{N-1} L_{N-1})}\left(\tau\right)\right] \cdot \\
            &\left[U_Z^{(1_{N-1} 2_{N-1} 3_{N-1} \cdots {L-1}_{N-1} L_{N-2})}\left(\tau\right)^\dagger U_{DQ}\left(\frac{t_{L-1}}{N}\right)  U_Z^{(1_{N-1} 2_{N-1} 3_{N-1} \cdots {L-1}_{N-1} L_{N-2})}\left(\tau\right)\right] \cdot \\
            & \qquad \qquad \qquad \vdots \\
            &\left[U_Z^{(1_{N-2} 2_{N-2} 3_{N-2} \cdots {L-1}_{N-2} L_{N-2})}\left(\tau\right)^\dagger U_{DQ}\left(\frac{t_L}{N}\right)  U_Z^{(1_{N-2} 2_{N-2} 3_{N-2} \cdots {L-1}_{N-2} L_{N-2})}\left(\tau\right)\right] \cdot \\
            & \qquad \qquad \qquad \vdots \\
            & \left[U_Z^{(1_1 2_1 3_1 \cdots {L-1}_1 L_1)}\left(\tau\right)^\dagger U_{DQ}\left(\frac{t_L}{N}\right)  U_Z^{(1_1 2_1 3_1 \cdots {L-1}_1 L_1)}\left(\tau\right)\right] \cdot \\
            & \qquad \qquad \qquad \vdots \\
            & \left[U_Z^{(1_1 2_0 3_0 \cdots {L-1}_0 L_0)}\left(\tau\right)^\dagger U_{DQ}\left(\frac{t_1}{N}\right)  U_Z^{(1_1 2_0 3_0 \cdots {L-1}_0 L_0)}\left(\tau\right)\right] \\
            \end{split} &&
        \end{flalign}
In terms of the toggling frame Hamiltonian, this can be written as
\begin{flalign}
        \begin{split}
            U_N = & \text{ exp}\left[-i \frac{t_L}{N} \mathcal{H}_m^{(1_N 2_N \cdots L_N)}\right] \cdot \text{exp}\left[-i \frac{t_{L-1}}{N} \mathcal{H}_m^{(1_N 2_N \cdots L_{N-1})}\right] \cdots \text{exp}\left[-i \frac{t_1}{N} \mathcal{H}_m^{(1_N 2_{N-1} \cdots L_{N-1})}\right] \cdot \\
            & \text{ exp}\left[-i \frac{t_L}{N} \mathcal{H}_m^{(1_{N-1} 2_{N-1} \cdots L_{N-2})}\right] \cdot \text{exp}\left[-i \frac{t_{L-1}}{N} \mathcal{H}_m^{(1_{N-1} 2_{N-1} \cdots L_{N-2})}\right] \cdots \text{exp}\left[-i \frac{t_1}{N} \mathcal{H}_m^{(1_{N-1} 2_{N-2} \cdots L_{N-2})}\right] \cdot \\
            & \qquad \qquad \qquad \vdots \\
            & \text{ exp}\left[-i \frac{t_L}{N} \mathcal{H}_m^{(1_1 2_1 \cdots L_1)}\right] \cdot \text{exp}\left[-i \frac{t_{L-1}}{N} \mathcal{H}_m^{(1_1 2_1 \cdots L_0)}\right] \cdots \text{exp}\left[-i \frac{t_1}{N} \mathcal{H}_m^{(1_1 2_0 \cdots L_0)}\right]
        \end{split} &&
        \end{flalign}
The first-order average Hamiltonian in this case is given by
\begin{flalign}
\bar{\mathcal{H}} = \frac{1}{N\left(t_1+t_2+\cdots+t_L\right)}\left[t_1\mathcal{H}_1 + t_2\mathcal{H}_2 + \cdots t_L\mathcal{H}_L\right]
&&
\end{flalign}
In $L=2$ case, there were $t_1$ and $t_2$ series for central-peripheral and peripheral-peripheral cases. In the general case, there are $t_1$, $t_2$,...,$t_L$ series. Consider the $t_1$ series first. For central-peripheral, it is given by
\begin{flalign}
\begin{split}
S_1^{CP} =& \; \text{exp}\left[i\tau(\Omega_1+\omega_1)\right] + \\
          & \; \text{exp}\left[i\tau(2\Omega_1+\Omega_2+\cdots \Omega_L + 2\omega_1 + \omega_2 + \cdots + \omega_L)\right] + \\
          & \; \text{exp}\left[i\tau(3\Omega_1 + 2\Omega_2 + \cdots 2\Omega_L + 3\omega_1 + 2\omega_2 + \cdots + 2\omega_L)\right] + \\ 
        & \qquad \qquad \qquad \vdots \\
          & \; \text{exp}\left[i\tau(N\Omega_1 + (N-1)\Omega_2 + \cdots (N-1)\Omega_L + N\omega_1 + (N-1)\omega_2 + \cdots + (N-1)\omega_L)\right]        
\end{split}
&&
\end{flalign}
Denoting $\sum_i\Omega_i = \Omega$ and $\sum_i\omega_i=\omega$, $S_1^{CP}$ can be written as
\begin{flalign}
S_1^{CP} = \; \text{exp}\left[i\tau(\Omega_1+\omega_1)\right]\left\{1+\text{exp}\left[i\tau(\Omega+\omega)\right]+\text{exp}\left[2i\tau(\Omega+\omega)\right]+\cdots+\text{exp}\left[(N-1)i\tau(\Omega+\omega)\right]\right\}
&&
\end{flalign}
In terms of the filter function, this can be written as
\begin{flalign}
S_1^{CP} = \; \text{exp}\left[i\tau(\Omega_1+\omega_1)\right] \mathcal{F}_N\left(\Omega + \omega\right)
&&
\end{flalign}
Similarly, the $t_1$ series for peripheral-peripheral interactions is given by
\begin{flalign}
S_1^{PP} = \; \text{exp}\left[i\tau(2\omega_1)\right] \mathcal{F}_N\left(2\omega\right)
&&
\end{flalign}
We can similarly compute the remaining series for both central-central and peripheral-peripheral interactions. One can verify that the filter function part of the series remains the same. Only changing factor is the exponential term preceding it. The table below gives the argument of the exponential term (It gives $x$ where the exponential term is $\text{exp}\left(i\tau x\right)$).
  \begin{table}[H]
\centering
\caption{Multiplicative Term for Different Interactions}
\label{mult_table}
\begin{tabular}{c|c|c|}
\cline{2-3}
                                      & \textbf{Central-Peripheral}                             & \textbf{Peripheral-Peripheral}                          \\ \hline
\multicolumn{1}{|c|}{$S_1$}     & $\Omega_1 + \omega_1$                                   & $2\omega_1$                                   \\ \hline
\multicolumn{1}{|c|}{$S_2$}     & $\Omega_1 + \Omega_2 + \omega_1 + \omega_2$             & $2\omega_1 + 2\omega_2$             \\ \hline
\multicolumn{1}{|c|}{$S_3$}     & $\Omega_1 + \Omega_2 + \Omega_3 + \omega_1 + \omega_2 + \omega_3$           & $2\omega_1 + 2\omega_2 + 2\omega_3$           \\ \hline
\multicolumn{1}{|c|}{\vdots}          & \vdots                                                  & \vdots                                                  \\ \hline
\multicolumn{1}{|c|}{$S_{L-1}$} & $\Omega_1+\Omega_2+\cdots+\Omega_{L-1} + \omega_1+\omega_2+\cdots+\omega_{L-1}$ & $2\omega_1+2\omega_2+\cdots+2\omega_{L-1}$ \\ \hline
\multicolumn{1}{|c|}{$S_L$} & $\Omega_1+\Omega_2+\cdots+\Omega_{L-1}+\Omega_L + \omega_1+\omega_2+\cdots+\omega_{L-1}+\omega_L$ & $2\omega_1+2\omega_2+\cdots+2\omega_{L-1}+2\omega_L$ \\ \hline
\end{tabular}
\end{table}
\noindent
To decouple the peripheral-peripheral interactions, the condition obtained from the argument of the filter function is
\begin{flalign}
2\omega=(2l+1)\pi
&&
\end{flalign}
Similarly to retain the central-peripheral interactions, the condition is
\begin{flalign}
\Omega+\omega=2m\pi
&&
\end{flalign}
While these conditions decouple the peripheral-peripheral interactions, the strength of the central-peripheral interactions  will have an extra multiplicative factor. The average Hamiltonian is given by
\begin{flalign}
\begin{split}
\bar{\mathcal{H}} = \frac{\mathcal{H}_{DQ}^R}{N\left(t_1+t_2+\cdots+t_L\right)}\Bigl\{ &t_1Ne^{i\tau(\Omega_1+\omega_1)} + \\
& t_2Ne^{i\tau(\Omega_1+\Omega_2+\omega_1+\omega_2)} + \\
& t_3Ne^{i\tau(\Omega_1+\Omega_2+\Omega_3+\omega_1+\omega_2+\omega_3)} + \\
& \qquad \qquad \vdots \\
& t_1Ne^{i\tau(\Omega+\omega)}\Bigr\}
\end{split}
&&
\end{flalign}
We can impose additional conditions on the parameters analogous to the way we did in Equation \ref{eqn:extra_cond} so that the average Hamiltonian reduces to DQ Hamiltonian with only radial interactions $(\mathcal{H}_{DQ}^R)$. Setting each individual exponential term in above equation to 1 gives the general conditions given in Equation \ref{star_generalcond}. They also remove the dependence on time since the time coefficient gets canceled with the total time in the denominator of the average Hamiltonian.
\end{document}